\documentclass[11pt]{article}
\usepackage{amsmath,amssymb,amsthm}
\usepackage{graphicx}
\usepackage{hyperref}
\usepackage{geometry}
\usepackage{amsmath,amssymb}
\usepackage{graphicx}
\usepackage{subcaption}
\usepackage{tikz}
\usetikzlibrary{arrows.meta, automata, positioning}
\usepackage{xspace}
\usepackage{booktabs}
\usepackage{multirow}
\usepackage{enumitem}
\usepackage{float}
\usepackage{algorithm}
\usepackage{algpseudocode}

\usepackage{hyperref}
\usepackage{tikz}
\usepackage{pgfplots}
\pgfplotsset{compat=1.18}

\usepackage{cite}

\date{} 

\title{\Large \bf Privacy-Preserving Covert Communication Using Encrypted Wearable Gesture Recognition}

\author{
Tasnia Ashrafi Heya \\
Department of Computer Science \\
University of Dayton \\
\texttt{tashrafiheya1@udayton.edu}
\and
Sayed Erfan Arefin \\
Department of Computer Science \\
University of Dayton \\
\texttt{sarefin1@udayton.edu}
}

\begin{document}
\maketitle

\begin{abstract}
Secure communication is essential in covert and safety-critical settings where verbal interactions may expose user intent or operational context. Wearable gesture-based communication enables low-effort, nonverbal interaction, but existing systems leak motion data, intermediate representations, or inference outputs to untrusted infrastructure, enabling intent inference, behavioral biometric leakage, and insider attacks. This work proposes a privacy-preserving gesture-based covert communication system that ensures, no raw sensor signals, learned features, or classification outputs are exposed to any third-party. The system employs a multi-party homomorphic learning pipeline for gesture recognition directly over encrypted motion data, preventing adversaries from inferring gesture semantics, replaying sensor traces, or accessing intermediate representations. To our knowledge, this work is the first to apply encrypted gesture recognition in a wearable-based covert communication setting. We design and evaluate haptic and visual feedback mechanisms for covert signal delivery and evaluate the system using 600 gesture samples from a commodity smartwatch, achieving over 94.44\% classification accuracy and demonstrating the feasibility of the proposed system with practical deployability from high-performance systems to resource-constrained edge devices.



\end{abstract}

\section{Introduction}
\vspace{-0.2cm}
Security-sensitive operations increasingly rely on \emph{covert}, low-bandwidth communication channels to coordinate actions without revealing intent, identity, or operational context. In many real-world scenarios—including personal safety, assistive technologies, and tactical coordination—traditional verbal or visual communication channels are unavailable or actively dangerous. Wearable gesture-based communication offers an appealing alternative: gestures are fast, discreet, and require minimal cognitive or motor effort. However, most existing gesture-based systems are not designed for adversarial environments.

Traditional wearable gesture systems routinely transmit raw or lightly processed motion data—such as accelerometer and gyroscope streams, intermediate features, or classification outputs—through smartphones, wireless links, cloud services, or edge servers that are honest-but-curious at best and actively compromised at worst. Prior work has shown that even low-level motion signals can leak sensitive information beyond their intended use, including behavioral patterns, biometric traits, health conditions, and personal attributes~\cite{watchyourwatch2023,motionspy2019}. These risks are exacerbated when gesture pipelines decrypt data at intermediate stages, exposing features or outputs to network adversaries, compromised devices, or insider operators. Consequently, the attack surface includes interception and inference during transmission~\cite{watchyourwatch2023}, spoofing and replay of recorded motion traces~\cite{replaybiometric2017,replaybiometric2018}, server-side decryption enabling insider threats~\cite{serverside1}, unauthorized profiling~\cite{wearableprivacy,emo4,speech1}, training-time poisoning or backdoors~\cite{backdoor,poisopning}, and adversarial or surveillance attacks on wearable datasets~\cite{wearableprivacy,adversarial,wearableprivacy2}.

These threats disproportionately affect users who depend on nonverbal communication and cannot easily switch modalities to mitigate exposure. In covert settings, privacy failures directly impact safety, autonomy, and trust. Addressing these risks requires treating gesture-based communication as a \emph{systems security problem}, rather than an application-layer concern. This, in turn, demands machine learning systems that can train and operate on protected data without exposing sensitive information to third parties, enforcing privacy as a first-class constraint.

This work addresses the following question: \emph{Can wearable gesture-based communication remain usable and responsive while providing strong protection against inference by untrusted infrastructure?} Protecting only raw sensor data is insufficient; intermediate representations and model outputs are equally sensitive and must be kept confidential. To this end, we propose a privacy-preserving gesture-based covert communication system that performs gesture recognition directly over encrypted wrist-based motion data using a multi-party homomorphic learning pipeline on commodity wearables. No raw signals, features, or classification outputs are ever revealed to third-party infrastructure.

Unlike prior encrypted inference efforts that focus on generic benchmarks or static datasets, our system targets interactive, real-time covert communication and explicitly considers adversaries that observe network traffic, stored ciphertexts, and repeated gesture executions. Training of motion sensing features is therefore performed using homomorphic multi-party machine learning in this work, ensuring that motion data, intermediate features, and model parameters are never exposed in plaintext. In contrast to Trusted Execution Environments (TEEs), which rely on hardware trust assumptions and remain vulnerable to privileged insiders and side-channel attacks~\cite{sgx_attack,sgx_attack2}, our HE+MPC design provides cryptographic isolation by construction, enabling secure operation across organizational trust boundaries without trusting hardware vendors or service operators.

The system integrates three components into a unified architecture: (1) wearable sensing with secure communication over untrusted networks, (2) privacy-preserving learning and inference using a multi-party homomorphic pipeline, and (3) covert signal delivery via receiver-side feedback. To support practical deployment, we design and study two retrieval mechanisms—a \emph{haptic feedback approach} and a \emph{visual feedback approach}—that deliver inferred intents discreetly while preserving accessibility. We evaluate the system using motion data collected from nine users wearing a commodity smartwatch, yielding 600 gesture instances. Despite performing classification entirely on encrypted data, the system achieves over 92\% accuracy under real-time constraints, demonstrating end-to-end protection against inference, replay, and unauthorized access while preserving responsive gesture recognition.

This work makes the following contributions:
\begin{itemize}
    \item \textbf{Encrypted gesture recognition pipeline:} We design and implement a multi-party homomorphic gesture recognition system that performs inference directly over encrypted wearable motion features, preventing leakage of raw signals, intermediate representations, and outputs.
    
    \item \textbf{Covert and accessible signal delivery:} We design and evaluate haptic and visual feedback mechanisms suitable for covert communication and assistive use under adversarial observation.
    
    \item \textbf{End-to-end practical evaluation:} We demonstrate feasibility through an end-to-end implementation on commodity smartwatches, achieving over 94\% classification accuracy under encrypted inference while maintaining interactive performance.

    \item \textbf{Edge-aware deployment evaluation:} We evaluate the system across heterogeneous execution environments, including NVIDIA Jetson Nano and Jetson Orin edge platforms as well as traditional server systems, to characterize performance and feasibility under realistic edge-computing constraints.
\end{itemize}

\section{Related Work}
\vspace{-0.3cm}
Our proposed work closely aligns with three different lines of past research: 1) hand gesture recognition with smartwatch, 2) privacy-preserving techniques, and 3) covert communication techniques.

\subsection{Hand Gesture Recognition Techniques}
\vspace{-0.2cm}
Extensive prior research has demonstrated that gestures performed in the air \cite{moazen2016airdraw, arduser2016recognizing}, on a whiteboard \cite{wijewickrama2019dewristified}, or on paper \cite{moazen2016airdraw, xia2018motionhacker} can be accurately recognized using biometric data collected from smartwatches. These studies explore various applications, including classroom transcription and potential side-channel attacks, where the technique could infer content typed or written by individuals wearing a smartwatch. Authors of \cite{mpoor_2} have examined user preference and ergonomics in touchless gestural manipulation, identifying gesture designs that minimize fatigue and improve usability. In subsequent research \cite{mpoor_1}, the authors investigated gesture-based interaction for users with chronic pain, demonstrating that modeling user-specific interaction spaces and reducing fine motor demands can significantly improve comfort and performance. In a closely related work \cite{watch_paper}, the authors demonstrated that smartwatch motion sensor data can reliably support wrist-based alphanumeric gesture recognition using lightweight machine learning models, providing a foundation for modeling fine-grained wrist motion and designing low-effort gesture vocabularies.

Existing gesture-based communication systems largely assume trusted execution environments or benign infrastructure and do not address adversarial inference, replay, or insider threats in covert settings.
\vspace{-0.2cm}

\subsection{Privacy-preserving Techniques}
\vspace{-0.15cm}
\subsubsection{Homomorphic Machine Learning}
\vspace{-0.15cm}
Homomorphic encryption is a key technique for privacy-preserving machine learning as well as multi-party computations (MPC). Gupta et al. \cite{gupta} used homomorphic filtering with K-means clustering for PCG signals, ensuring privacy in feature extraction. As Machine Learning as a Service (MLaaS) grows, methods like combining homomorphic encryption with neural networks \cite{cloud} enable encrypted data processing. CryptoNets \cite{cryptonet} demonstrated this approach with the MNIST dataset, showing secure neural network processing on encrypted data. Recent advances include multi-party protocols using Paillier encryption \cite{npmml} and decentralized deep learning frameworks like D2-MHE \cite{d2mhe} for secure, collaborative training. In another research \cite{MPC_HE} paper, the authors implemented and evaluated a multi-party homomorphic neural network (HNN) model in time-series unstructured data for smart VA applications. 

Our work differs fundamentally in scope and objective. Rather than treating encrypted inference as an abstract machine learning primitive, we apply homomorphic learning to a \emph{real-time, interactive wearable system} operating in adversarial environments. 

\subsubsection{Privacy-preserving Gesture Recognition}
In a closely related work with us is \cite{PP_ges1} where Ignasiak et al. propose a cloud-trained gesture recognition system that protects user privacy by transmitting only locally extracted and permuted hand-landmark features and train the features with standard neural network training under an honest-but-curious cloud assumption. Their obfuscation-based approach primarily focuses on camera-based hand gesture datasets, including ASL and live webcam captures. In another work closely align with our work is \cite{PP_ges2}, where Smedemark-Margulies et al. propose a fast and expressive EMG-based gesture recognition system that learns a combination-homomorphic feature space, enabling generalization from single gestures to unseen gesture combinations using contrastive pretraining and synthetic data augmentation, evaluated on a supervised multi-subject sEMG dataset collected with forearm electrodes. 

In contrast, our work performs gesture recognition directly over encrypted wearable motion sensor features using multi-party homomorphic learning, providing cryptographic end-to-end privacy guarantees on commodity wearables while supporting accessible, covert communication under stronger adversarial models and without requiring supervised combination-gesture calibration or reliance on vision- or EMG-based datasets.

\subsection{Covert Communication Techniques}
\vspace{-0.2cm}
Covert communication exploits vulnerabilities across wireless, mobile, and wearable systems. In wireless networks, covert data transmission through relay nodes \cite{relay_paper1, relay_paper2} and deep learning models \cite{deep_learning_paper} enhance covert communication, making detection difficult. In mobile devices, Android applications often exhibit covert communication \cite{mobile_apps_paper}, while Android wearables use status bar notifications for hidden data channels \cite{wearable_paper}. Covert methods have also emerged in federated learning systems to secure data exchanges via friendly jammers \cite{federated_learning_paper}, demonstrating the broad need for enhanced security in these technologies.

In summary, this work bridges the gap between encrypted machine learning and practical wearable security. While prior encrypted ML research focuses on general-purpose inference under encryption, we demonstrate how homomorphic learning can be applied to secure \emph{covert gesture-based communication}, mitigating realistic adversaries. To our knowledge, this is the first system to integrate multi-party homomorphic gesture recognition into a wearable communication pipeline, providing end-to-end protection against inference, replay, and insider threats while remaining usable on commodity devices, under realistic edge constraints.


\section{Experimental Design}
\subsection{Design Requirements}
\label{require-1}

We identify three core design requirements that guide the design of our multi-party homomorphic gesture-based covert communication system. These requirements reflect the need to support secure, discreet, and reliable nonverbal interaction under adversarial observation and untrusted infrastructure.

\textbf{Stealth and Observational Plausibility.}
A fundamental requirement for covert communication is that gesture generation and signal retrieval remain indistinguishable from benign user behavior. On the sender side, gestures must be naturally embedded within everyday wrist motion, such as writing or routine hand movement. We achieve this by encoding messages using subtle wrist gestures interleaved with short pauses, captured by inertial sensors on commodity smartwatches. These gestures blend with ordinary motion patterns and avoid exaggerated or attention-drawing movements. Motion data are transmitted to a paired mobile device and processed entirely under encryption, ensuring that covert signaling remains both behaviorally and cryptographically unobtrusive.

On the receiver side, signal retrieval must also avoid drawing attention. To this end, we design two retrieval mechanisms: a \emph{visual approach}, which conveys decoded signals through unobtrusive on-screen cues, and a \emph{haptic approach}, which encodes information using vibration patterns. These modalities allow recipients to retrieve messages without overt interaction and may be used independently or in combination to expand the message space while preserving discretion.

\textbf{Recognition Accuracy under Encryption.}
Covert communication is effective only if gesture signals are recognized reliably. The system must accurately distinguish among gesture classes despite inter-user variability, sensor noise, and the computational constraints imposed by encrypted execution. Prior work shows that wrist-motion patterns vary significantly across users, degrading population-wide models. Our system therefore supports user-specific gesture modeling while performing both training and inference directly on encrypted representations using multi-party homomorphic machine learning. This design preserves recognition accuracy without exposing raw motion data, intermediate features, or model states to untrusted devices or servers.

\begin{figure*}
    \centering
    \includegraphics[width=\linewidth]{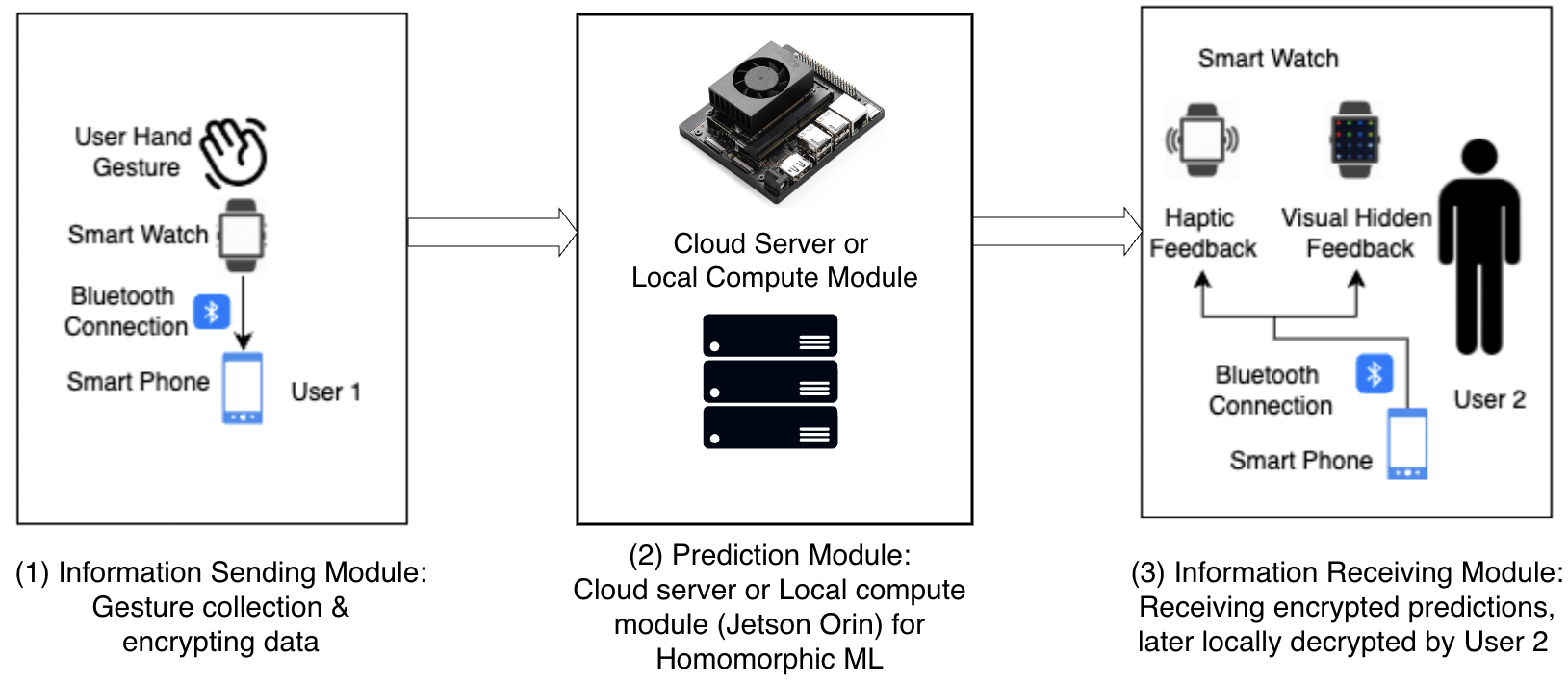}
    \caption{System overview.}
    \label{fig:system_overview}
\end{figure*}

\textbf{Ease of Execution and Low Cognitive Load.}
Finally, the system must be practical for real-world use. Gesture inputs should be easy to perform repeatedly and should not require fine motor precision or sustained cognitive effort. Complex or physically demanding gestures increase error rates and reduce usability, particularly in high-stress or assistive settings. We therefore restrict the gesture vocabulary to simple wrist motions and pauses that resemble natural writing or routine hand movements. This design supports stealth, recognition robustness, and user comfort, enabling prolonged use without training-intensive interaction. \emph{To further reduce cognitive and operational overhead, we evaluate encrypted gesture-recognition training and inference on resource-constrained edge platforms, including the NVIDIA Jetson Nano and Jetson Orin.} These results show that the system can operate locally under limited compute and power budgets, making it suitable for covert deployment without reliance on cloud infrastructure.

Overall, these requirements ensure that the system supports secure, covert, and usable nonverbal communication while operating entirely on encrypted data under accessible sensing and deployment constraints.

\subsection{System Design}

Figure~\ref{fig:system_overview} illustrates the end-to-end architecture of the proposed privacy-preserving gesture-based covert communication system. The system comprises three logical modules: (1) an \emph{Information Sending Module} for gesture capture and encryption, (2) a \emph{Prediction Module} that performs homomorphic inference over encrypted data, and (3) an \emph{Information Receiving Module} that decrypts and renders the inferred intent via covert feedback. Privacy is enforced by construction: raw sensor data, intermediate features, and inference outputs remain encrypted end-to-end and are never exposed in plaintext outside user-controlled devices.

\subsubsection{Information Sending Module}
\label{sec:design_sending}
User~1 performs a predefined wrist gesture using a commodity smartwatch. Inertial measurements are transmitted to a paired smartphone over Bluetooth, where lightweight preprocessing consistent with the trained classifier’s input representation is applied. The resulting feature vector is immediately encrypted under a receiver-authorized homomorphic encryption (HE) public key, ensuring that neither raw motion signals nor derived features leave the user’s device in plaintext.

\subsubsection{Prediction Module}
\label{sec:design_prediction}
The Prediction Module receives encrypted feature vectors and applies a homomorphic neural network (HNN) to compute encrypted predictions. Inference may execute on either a cloud server or a local edge device (e.g., Jetson Orin Nano), enabling flexible deployment without altering the privacy model. Because computation is performed entirely over ciphertexts, the module never observes raw inputs, intermediate activations, or predicted labels. The HNN is constructed using HE-compatible operations, allowing the same encrypted model to scale across heterogeneous hardware platforms.

\subsubsection{Information Receiving Module}
\label{sec:design_receiving}
Encrypted predictions are delivered to User~2’s smartwatch, where local decryption is performed using the corresponding private key. Only the intended recipient can recover the semantic output, which is conveyed through covert channels such as haptic feedback or visually hidden cues. This design ensures that intent disclosure remains limited to the authorized endpoint and unobtrusive to external observers.

Overall, the system enforces privacy by construction through end-to-end homomorphic encryption, structurally precluding plaintext observation by intermediate infrastructure. Privacy preservation follows directly from the cryptographic properties of the homomorphic scheme, allowing evaluation to focus on feasibility, accuracy, and performance rather than leakage mitigation.

\subsection{Data Collection}
\label{data-1}

\subsubsection{Exploratory Evaluation of Gesture Symbols}
To select a set of gesture symbols that balance usability and classification robustness, we conducted an exploratory evaluation of candidate wrist-motion patterns. While it is desirable for gesture symbols to be intuitive and easy to remember, reliable system operation additionally requires that their motion signatures be well separated under realistic sensing noise and inter-user variability. We therefore evaluated a range of alphanumeric symbols by analyzing their gyroscope-derived feature representations and measuring both classification accuracy and confusion patterns. Initial experiments with four user participant data showed that symbols corresponding to simple, visually distinct motion patterns (e.g., A, B, and C) consistently achieved high separability, whereas others (notably symbols with similar stroke structure, e.g., D, which is similar to B) were frequently confused and significantly degraded overall accuracy.

\begin{table}[ht] 
\caption{Results from clustering of gesture symbols.} 

\begin{center} 
\resizebox{0.7\textwidth}{!}{
\begin{tabular}{c|c|c|c|c} 
\hline 
\textbf{User No.} & \multicolumn{4}{|c}{\textbf{Data Clusters}} \\ \cline{2-5} \textbf{} & \textbf{\textit{Cluster 1}}& \textbf{\textit{Cluster 2}}& \textbf{\textit{Cluster 3}} & \textbf{\textit{Cluster 4}}\\ 
\hline \hline
1 & {\bf{B}},H,J,O,q,S,y,Z,8 & {\bf{C}},K,m & {\bf{E}},I,W & {\bf{A}},X \\ 

2 & {\bf{B}},I,J,m,O,q,S,8 & {\bf{C}},X & A,D,{\bf{E}},H,W,y,Z & K \\ 

3 & {\bf{B}},K,X,Z & {\bf{C}},J,m,O,q,S,W,y,8 & {\bf{E}},H,I & {\bf{A}} \\ 

4 & {\bf{B}},I,J,K,W,X & {\bf{C}},q & {\bf{E}},O,S,Z,8 & {\bf{A}},H,m,y \\ 
\hline 
\end{tabular} 
} 
\label{cluster-1} 
\end{center} 
\end{table} 

To systematically identify a robust symbol set, we combined visual filtering with unsupervised clustering. After discarding visually similar characters, we collected motion data for the remaining candidates and applied K-means clustering ($K=4$) over the extracted feature space. Table~\ref{cluster-1} summarizes representative clustering results across users. Three symbols (A, B, and C) consistently occupied distinct clusters, indicating strong separability across users. The remaining cluster was dominated by the symbol E, which exhibited stable separation from the others. Based on these results, we selected the final gesture symbol set $\{A, B, C, E\}$, which provides a compact, memorable vocabulary while maintaining high classification robustness. This symbol set is used throughout the full data collection and encrypted inference evaluation described in Section~\ref{data-1}.

\subsubsection{Experimental Data Collection}
With IRB approval, we collected wrist-motion data to evaluate our privacy-preserving gesture recognition under realistic sensing noise and inter-user variability. The data collection phase captures wrist-based motion sensing data which is used for feature extraction and the encrypted features then transerred to serve as input to gesture recognition and encrypted inference, with each gesture symbol mapped to a predefined semantic signal (e.g., alert or assistance request). The goal of data collection is not to build a large-scale biometric dataset, but to characterize gesture separability, segmentation reliability, and model robustness in the presence of natural human motion.

Data were collected from nine user participants wearing an Fossil Gen 6 smartwatch on their dominant wrist. Each participant performed 15 repetitions of four predefined gesture symbols (\textit{A, B, C, E}), yielding a total of 540 gesture instances. Gesture symbols were selected to require low motor precision and minimal cognitive load, reflecting realistic use in assistive and safety-critical settings. The smartwatch recorded tri-axial gyroscope and accelerometer measurements at 60~Hz. Sensor data were transmitted via Bluetooth to a paired smartphone, where they were logged for subsequent preprocessing and analysis. 

Participants were instructed to perform gestures naturally, allowing variation in speed, intensity, and wrist posture. To collect data, the user pretended to write the letters (e.g., letter ``A'') immediately after the press of a start button on the app interface. On completion of writing the letters, a stop button was immediately pressed so gyroscope data that precisely delimits these letters was saved on the mobile device. This process was repeated 8 more times for each user participants. The collected traces include both gesture motion segments and intervening pause regions, enabling evaluation of the temporal segmentation protocol described in Section~\ref{sec:gesture-encoding}. In particular, it supports analysis of inter-user variability, gesture boundary detection, and classification accuracy under homomorphic inference. All reported results are derived from this dataset, which forms the basis for evaluating system performance in both plaintext and encrypted execution modes.

\subsection{Gesture Encoding and Temporal Segmentation}
\label{sec:gesture-encoding}

The system encodes covert messages using a finite set of \emph{gesture symbols} derived from wrist-motion patterns captured by inertial sensors on commodity wearable devices. Each symbol maps to a predefined semantic signal (e.g., alert, abort, request for assistance), and symbol sequences enable compound messages. Gesture symbols are designed to resemble everyday wrist motion, ensuring behavioral indistinguishability from benign activity under visual observation.

\begin{figure}[t]
\centering
\includegraphics[width=0.73\textwidth]{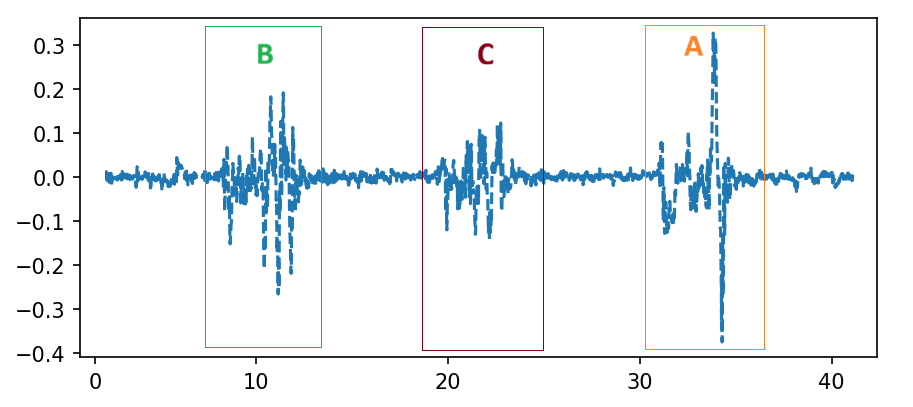}
\caption{Example of temporal segmentation on a continuous wrist-motion stream. Highlighted windows indicate detected gesture-symbol segments (e.g., B, C, A), separated by pause.}
\label{fig:segmentation_example}
\end{figure}

Gesture instances are identified using a \emph{temporal segmentation protocol} that alternates between motion and pause phases. Pauses act as implicit delimiters between gesture symbols, eliminating the need for explicit user actions or visible cues. An \emph{opening pause} transitions the system from an idle monitoring state to an active communication state, motion segments correspond to gesture symbols, and a \emph{closing pause} terminates the communication window. Figure~\ref{fig:segmentation_example} illustrates an end-to-end example of this protocol on a continuous wrist-motion trace, where consecutive gesture symbols are separated by pause regions and segmented into symbol windows.

\begin{figure}[t]
\centering
\begin{tikzpicture}[
  >=Stealth,
  node distance=1.48cm,
  every state/.style={
    draw, rectangle, rounded corners,
    minimum height=0.95cm,
    text width=1.68cm,
    align=center
  }
]

\node[state] (idle)   {Idle\\monitor};
\node[state, right=of idle] (active) {Active\\encrypt + infer};
\node[state, right=of active] (closed) {Closed\\reset};

\draw[->] (idle) -- (active)
  node[pos=0.5, above, align=center]{Opening\\pause}
  node[pos=0.5, below, font=\footnotesize] {$t\!\in\!(t_1\!\pm\!\varepsilon)$};

\draw[->] (active) edge[loop above, looseness=6]
  node[above, font=\footnotesize, yshift=1pt] {Gesture symbol}
  (active);

\draw[->] (active) -- (closed)
  node[pos=0.5, above, align=center]{Closing\\pause}
  node[pos=0.5, below, font=\footnotesize] {$t\!\in\!(t_2\!\pm\!\varepsilon)$};

\draw[->] (closed.south) .. controls +(0,-0.75) and +(0,-0.75) .. (idle.south)
  node[midway, below, font=\footnotesize] {Reset};

\end{tikzpicture}
\caption{State-machine model for temporally segmented gesture communication. Opening and closing pauses delimit the active inference window.
}
\label{fig:state_machine}
\end{figure}
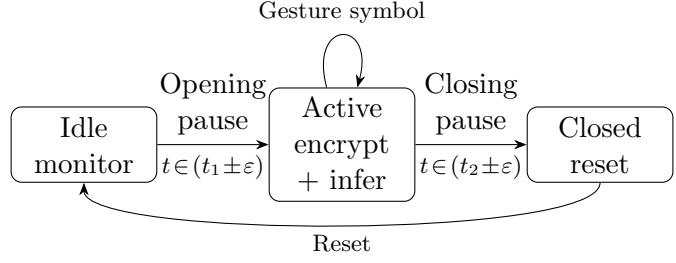

Gesture communication is modeled as a finite-state process (Figure~\ref{fig:state_machine}). The system begins in the \emph{idle} state, where motion is observed but ignored. Detection of an opening pause triggers a transition to the \emph{active} state, during which motion segments are buffered, encrypted on-device, and forwarded for privacy-preserving inference. Gesture symbols correspond to self-loops in the active state. Detection of a closing pause transitions the system to a \emph{closed (reset)} state, after which it returns to idle and ignores subsequent motion until reactivated.

\paragraph{Design Rationale.}
Pauses are used as segmentation cues because they are naturally occurring during human motion, robust to inter-user variability, and easy to detect without introducing explicit control signals that would reduce covertness. Discrete gesture symbols provide a compact and semantically meaningful representation that simplifies encrypted inference and limits information leakage compared to raw continuous streams. Continuous inference over unsegmented motion was avoided because it increases computational cost under homomorphic evaluation and exposes timing and length side channels, whereas pause-delimited windows bound both inference complexity and observable metadata.

\begin{figure*}[t]
    \centering
    \begin{subfigure}[t]{0.32\textwidth}
        \centering
        \includegraphics[width=\linewidth]{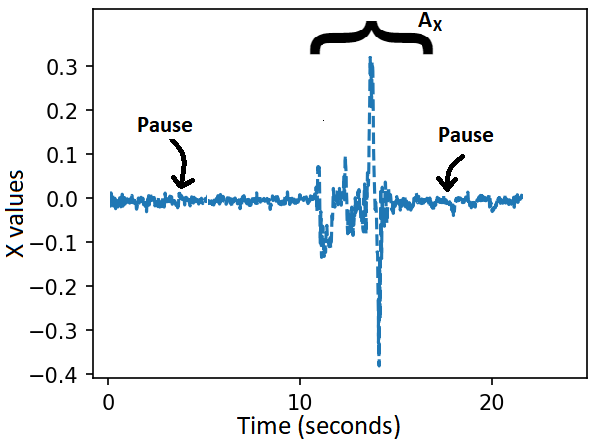}
        \caption{Gyroscope X-axis.}
    \end{subfigure}\hfill
    \begin{subfigure}[t]{0.32\textwidth}
        \centering
        \includegraphics[width=\linewidth]{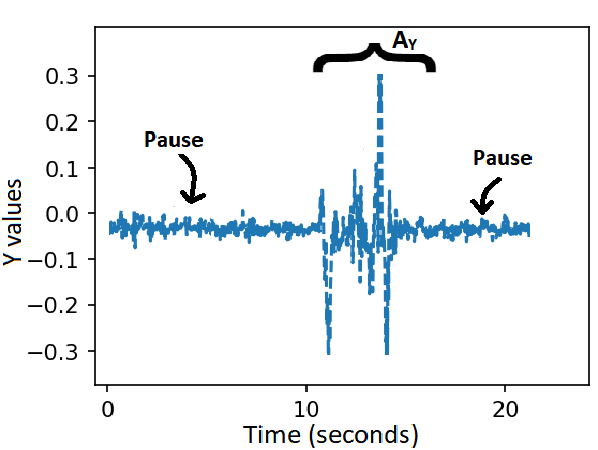}
        \caption{Gyroscope Y-axis.}
    \end{subfigure}\hfill
    \begin{subfigure}[t]{0.32\textwidth}
        \centering
        \includegraphics[width=\linewidth]{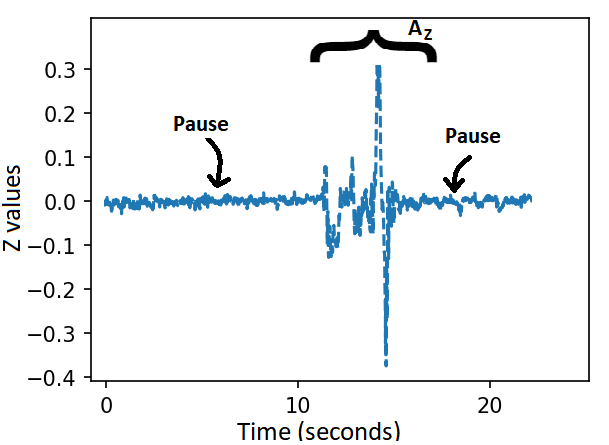}
        \caption{Gyroscope Z-axis.}
    \end{subfigure}

    \caption{Example wrist-motion gyroscope traces illustrating the temporal segmentation protocol.
    Extended pauses indicate gesture boundaries, while motion segments correspond to individual gesture symbols.}
    \label{fig:LetterSeparation}
\end{figure*}

\paragraph{Illustrative Action Sequences.}
Figure~\ref{fig:state_machine} also provides an operational interpretation of gesture-based communication. Assuming the sender transmits a single gesture-based signal, the system supports two equivalent action sequences that realize identical state transitions.

\textbf{Action Sequence 1: Pause-Based Activation.}
\begin{enumerate}
    \item A pause of $(t_1 \pm \varepsilon)$ seconds is detected as an opening pause, transitioning the system from \emph{idle} to \emph{active}.
    \item A wrist-motion pattern corresponding to a gesture symbol is executed and processed while the system remains in the \emph{active} state.
    \item A pause of $(t_2 \pm \varepsilon)$ seconds is detected as a closing pause, transitioning the system to \emph{closed} and then back to \emph{idle}.
\end{enumerate}

\textbf{Action Sequence 2: Symbol-Based Activation.}
\begin{enumerate}
    \item A predefined activation symbol (e.g., $\phi$ or $\Omega$), following a pause of $(t_1 \pm \varepsilon)$ seconds, transitions the system to the \emph{active} state.
    \item The intended gesture symbol is executed and processed while remaining in the \emph{active} state.
    \item A predefined termination symbol, following a pause of $(t_2 \pm \varepsilon)$ seconds, transitions the system to \emph{closed} and then to \emph{idle}.
\end{enumerate}

In both sequences, pauses function solely as temporal delimiters, and all motion data, intermediate features, and inference outputs remain encrypted throughout processing. An external observer without decryption capability cannot distinguish gesture boundaries, infer message length, or recover gesture semantics. All segmentation, feature extraction, and classification operate on encrypted representations using multi-party homomorphic machine learning. Consequently, raw motion data, intermediate features, and model states are never exposed in plaintext outside the user device. Figure~\ref{fig:LetterSeparation} illustrates representative gyroscope traces across all three axes, highlighting the separation between motion segments and temporal pauses that enables reliable segmentation under noisy motion and inter-user variability.

The protocol supports both single- and multi-symbol messages without modification. After termination, the system returns to the idle state and ignores subsequent motion, preventing unintended signal leakage. By abstracting gesture communication as a state-driven, temporally segmented process operating entirely on encrypted data, the system enables covert, reliable, and adversary-resilient nonverbal communication under realistic sensing and deployment constraints.


\subsection{Feature Extraction}
\label{sec:feature-extraction}

Feature extraction operates on temporally segmented wrist-motion traces produced by the pause-based segmentation protocol described in Section~\ref{sec:gesture-encoding}.

\paragraph{Pause Detection and Gesture Windowing.}
Pause detection is performed by monitoring instantaneous gyroscope measurements along each axis. A sample $g_x$ is flagged as a candidate pause point if it satisfies $-T_h \leq g_x \leq T_h$, where $T_h$ is a specific threshold calibrated from training data containing representative pause segments. Consecutive samples meeting this condition are grouped and evaluated against duration constraints to determine whether they correspond to valid opening or closing pauses. Samples that do not meet these criteria are ignored. Once a valid opening--closing pause pair is identified, all gyroscope samples between them are treated as a single gesture instance. This windowing step isolates gesture motion from surrounding background activity and ensures consistent downstream feature extraction.

\paragraph{Feature Computation.}
For each segmented gesture window, features are extracted independently from the $X$, $Y$, and $Z$ axes of the gyroscope signal. For each axis, the system computes 32 features spanning both temporal and spectral domains, resulting in a 96-dimensional feature vector per gesture instance. Temporal features capture amplitude, variability, distributional shape, and structural properties of the wrist-motion signal in the time domain. These features are effective at characterizing motion intensity, smoothness, and consistency across repetitions. Spectral features summarize the frequency content and energy distribution of the signal, capturing differences in gesture execution speed and oscillatory behavior that are not apparent in the time domain alone. Table~\ref{features-1} lists all temporal and spectral features used in our experiments.

The extracted feature vectors are used to train and evaluate gesture classifiers. The same feature representation is used for both unencrypted and encrypted gesture recognition evaluation, enabling direct comparison of performance across execution modes. For the encrypted inference evaluation, the dataset with extracted features are encrypted as cryptensors and transferred to the training module.

\begin{table*}[ht]
\caption{List of temporal and spectral features used to train our classifiers}
\begin{center}
\resizebox{\textwidth}{!}{%
\begin{tabular}{c|c}
\hline
\textbf{Type} & \textbf{Feature Names} \\
\hline \hline
Temporal features &
Mean, Standard Deviation, Interquartile Range, Absolute Energy, Mean Absolute Deviation, \\
& Standard Error of the Mean, Mean Change, Autocovariance, Longest Strike Above Mean, Variance, \\
& Absolute Sum of Changes, Kurtosis, Sample Entropy, Autocorrelation, Mean Absolute Change, \\
& Sum, Skewness, Quantile, Median, Longest Strike Below Mean, Complexity Invariant Distance \\
\hline
Spectral features &
Spectral Centroid, Spectral Flatness, Spectral Kurtosis, Spectral Skewness, Spectral Decrease, \\
& Spectral Spread, Spectral Rolloff, Spectral Slope \\
\hline
\end{tabular}
}
\label{features-1}
\end{center}
\end{table*}

\subsection{Theory of Multi-Party Encryption}
Our system implements privacy-preserving arithmetic operations—including addition, subtraction, multiplication, division, matrix operations, and limited nonlinear transformations—using the CrypTen library~\cite{crypten}, a framework for secure multi-party computation (MPC) developed by Meta. CrypTen provides a high-level abstraction for encrypted and secret-shared computation and interoperates with lattice-based cryptographic backends such as HElib and Microsoft SEAL~\cite{seal,helib}. Together, these components enable arithmetic over protected values without exposing plaintext data to any participating party. CrypTen combines MPC protocols with leveled homomorphic encryption (LHE) techniques to support machine learning workloads over encrypted data. Since most lattice-based schemes operate natively over integers, CrypTen employs fixed-point arithmetic to approximate real-valued computation, avoiding the complexity and inefficiency of directly supporting IEEE floating-point representations under homomorphic evaluation.

\subsubsection{Homomorphic Scheme}

CrypTen relies on lattice-based constructions derived from Brakerski–Gentry–Vaikuntanathan (BGV)-style homomorphic encryption schemes~\cite{bv,mpc_she}. These schemes fall under the category of leveled (or somewhat) homomorphic encryption and support a bounded number of additions and multiplications before ciphertext noise exceeds the tolerance required for correct decryption. The security of these lattice-based constructions is based on the Learning With Errors (LWE) problem, which is widely regarded as computationally hard even for quantum adversaries~\cite{bv,mpc_she}. Noise injected during encryption ensures semantic security but constrains the multiplicative depth of supported computations.

\textit{Key Generation.}
Given a lattice dimension $n$ and modulus $q$, the secret key is defined as $sk = (1, -s)$, where $s \sim \chi_{\text{key}}$ is sampled from a noise distribution. The public key is constructed as
\begin{equation}
    pk = (b, a) \in \mathbb{Z}_q^n \times \mathbb{Z}_q^n, \quad \text{with } b = a s + e \bmod q,
\end{equation}
where $a$ is sampled uniformly from $\mathbb{Z}_q^n$ and $e \sim \chi_{\text{err}}$ is drawn from an error distribution.

\textit{Encryption.}
A binary message $m \in \{0,1\}$ is encrypted as a ciphertext $c = (c_0, c_1)$:
\begin{align}
    u &\gets \chi_{\text{enc}}, \\
    e_1, e_2 &\gets \chi_{\text{err}}, \\
    c_0 &= b \cdot u + e_1 + 2m \bmod q, \\
    c_1 &= a \cdot u + e_2 \bmod q.
\end{align}

\textit{Decryption.}
Given $sk = (1, -s)$, decryption recovers the plaintext as
\[
Dec_{sk}(c) = \left\lfloor \frac{(c_0 + c_1 \cdot s) \bmod q}{q/2} \right\rfloor \bmod 2,
\]
which removes the secret-key-dependent term and tolerates bounded noise accumulated during homomorphic evaluation.

\subsubsection{Operations in CrypTen}

Rather than relying exclusively on pure homomorphic encryption, CrypTen primarily uses arithmetic secret sharing combined with MPC protocols. Secret-shared values are distributed across participating parties, enabling collaborative computation without revealing individual inputs or intermediate values. Interactive operations, such as secure multiplication, are accelerated using Beaver triples generated during an offline preprocessing phase or via a trusted setup~\cite{crypten,mpc_she}.

CrypTen further integrates high-performance communication backends, including Gloo and NCCL, and supports GPU acceleration through optimized CUDA libraries such as cuBLAS and cuDNN. This design enables scalable encrypted training and inference with practical performance overheads suitable for real-world machine learning workloads.

\subsubsection{Fixed-Point Encoding}

To support approximate real-valued computation over integer-based cryptographic primitives, CrypTen represents real numbers using fixed-point encoding. Given a precision parameter $t$, a real value $x \in \mathbb{R}$ is encoded as
\[
x_{\text{scaled}} = \left\lfloor x \cdot 2^t \right\rceil,
\]
where $\left\lfloor \cdot \right\rceil$ denotes rounding to the nearest integer. All computations are performed over these scaled integers within a modular ring $\mathbb{Z}_n$, with the modulus $n$ chosen to prevent overflow during intermediate operations. After encrypted computation, results are decoded by dividing by $2^t$, yielding approximate real-valued outputs with bounded numerical error.

\subsection{Homomorphic Neural Network Design}
\label{sec:hnn-design}

To support privacy-preserving gesture recognition, we design a homomorphic neural network (HNN) that operates on encrypted motion features using secure multi-party computation (MPC). The model follows a conventional feed-forward neural network structure but is implemented using cryptographic tensor abstractions to ensure that raw features, intermediate activations, and model parameters are never revealed in plaintext during training or inference.

We first implement and validate a plaintext neural network to establish baseline performance and guide architectural choices, including layer depth, activation functions, and loss formulation. We then construct an equivalent encrypted model using CrypTen-style secret-shared tensors and MPC protocols, ensuring that the forward pass, loss computation, and gradient updates are compatible with leveled homomorphic and MPC-based computation constraints. Algorithm~\ref{alg:hnn} summarizes the high-level training procedure. Unlike standard neural network training, encrypted training must carefully manage numerical precision, multiplicative depth, and interactive protocol costs. As a result, we evaluate multiple architectural configurations to mitigate overflow, noise growth, and accuracy degradation caused by repeated encrypted arithmetic operations.

\begin{algorithm}
\caption{Multi-Party Homomorphic Gesture Classification}
\label{alg:hnn}
\begin{algorithmic}[1]
\State \textbf{Input:} Encrypted gesture features $X_{\text{enc}}$, encrypted labels $Y_{\text{enc}}$, number of epochs $N$
\State \textbf{Output:} Decrypted gesture predictions at authorized endpoints
\State Initialize MPC context and secret-shared parameters
\State Initialize neural network parameters $\Phi$
\For{$i = 1$ to $N$}
    \State $Z_{\text{enc}} \gets \text{Forward}(X_{\text{enc}}, \Phi)$
    \State Compute encrypted loss $L_{\text{enc}}(Z_{\text{enc}}, Y_{\text{enc}})$
    \State Compute encrypted gradients $\nabla_{\Phi} L_{\text{enc}}$
    \State Securely update parameters $\Phi \leftarrow \Phi - \eta \nabla_{\Phi} L_{\text{enc}}$
\EndFor
\end{algorithmic}
\end{algorithm}

\subsubsection{Model Architecture and Initialization}

The homomorphic neural network is a fully connected classifier designed for \textbf{four-class gesture recognition}. Based on empirical evaluation under encrypted computation constraints, we adopt a compact three-layer architecture that balances classification accuracy with encrypted computation cost:
\[
\textbf{fc1: } \text{Linear}(d_{\text{in}}, 250),
\textbf{fc2: } \text{Linear}(250, 80),
\textbf{fc3: } \text{Linear}(80, 4).
\]

Here, $d_{\text{in}}$ denotes the dimensionality of the standardized motion feature vector extracted from wrist-sensor data. Deeper or wider architectures were evaluated but exhibited increased numerical instability, higher latency, and noise growth under encrypted training without commensurate accuracy gains. The selected architecture minimizes encrypted nonlinearity depth while preserving sufficient representational capacity for separable gesture features.

All weights are initialized from a scaled normal distribution,
\[
W \sim \mathcal{N}\!\left(0, \frac{1}{n_{\text{in}}}\right),
\]
where $n_{\text{in}}$ is the input dimension of the corresponding layer. Bias terms are initialized to zero. This initialization improves numerical stability and gradient propagation under fixed-point encrypted arithmetic. All parameters are represented as secret-shared encrypted tensors using CrypTen.

\subsubsection{Forward Pass}

Encrypted inputs are propagated through the network using linear transformations followed by Leaky ReLU activation functions. For each layer, the transformation is
\[
Y = XW + b,
\]
followed by element-wise activation. The final layer produces encrypted logits corresponding to the four gesture classes. Probability normalization (e.g., softmax) is deferred until decryption to avoid non-polynomial operations during encrypted evaluation.

\subsubsection{Activation Functions}

All intermediate layers employ the \textbf{Leaky Rectified Linear Unit (Leaky ReLU)} activation function. In our implementation, Leaky ReLU is realized via \emph{secure comparison-based MPC} rather than pure homomorphic evaluation. Specifically, encrypted comparisons and masked selection operations (e.g., \texttt{crypten.where}) are used to implement the piecewise function:
\[
\text{LeakyReLU}(x) =
\begin{cases}
x, & x > 0, \\
\alpha x, & x \leq 0,
\end{cases}
\]
where $\alpha$ is a small constant. This design avoids polynomial approximations while maintaining correct semantics under MPC. Compared to standard ReLU, Leaky ReLU preserves nonzero gradients for negative activations, improving gradient flow and numerical robustness under fixed-point encrypted backpropagation.

\subsubsection{Loss Computation}

The homomorphic model is trained using mean-squared error (MSE) loss, which avoids
non-polynomial operations such as logarithms during encrypted training. Given
encrypted logits $O \in \mathbb{R}^{N \times C}$ and one-hot encrypted labels
$Y \in \mathbb{R}^{N \times C}$ for $N$ samples and $C{=}4$ classes, our
implementation uses PyTorch-style \texttt{mean} reduction over all entries:
\[
L \;=\; \mathrm{mean}\!\left((Y - O)^2\right)
\;=\; \frac{1}{NC}\sum_{i=1}^{N}\sum_{c=1}^{C}(Y_{i,c}-O_{i,c})^2.
\]

In our implementation, the constant $1/C$ factor implied by the mean reduction
is absorbed into the learning rate~$\eta$; this does not affect convergence or correctness and simplifies encrypted gradient computation.

\subsubsection{Backward Pass and Parameter Updates}

The gradient with respect to the encrypted output logits is computed as
\[
\frac{\partial L}{\partial O} = \frac{2}{NC}(O - Y),
\]
where the constant $1/C$ factor is absorbed into the learning rate~$\eta$ in our implementation to simplify encrypted gradient computation. Backpropagation proceeds layer-by-layer under encryption. Given upstream gradient $G$, each layer computes
\[
\frac{\partial L}{\partial X} = GW^\top,\quad
\frac{\partial L}{\partial W} = X^\top G,\quad
\frac{\partial L}{\partial b} = \mathrm{mean}(G),
\]
where the mean is taken across the batch dimension. Parameters are updated using encrypted stochastic gradient descent:
\[
W \leftarrow W - \eta \frac{\partial L}{\partial W}, \quad
b \leftarrow b - \eta \frac{\partial L}{\partial b}.
\]

Between linear layers, gradients are gated by the encrypted derivative of Leaky ReLU, implemented via secure comparison and masking:
\[
\frac{\partial\,\mathrm{LeakyReLU}(z)}{\partial z} =
\begin{cases}
1, & z > 0,\\
\alpha, & z \le 0.
\end{cases}
\]
All intermediate activations, gradients, and parameter updates remain encrypted throughout training. Decryption is performed only for aggregate metrics (e.g., loss logging or final predictions) at authorized endpoints and does not affect the security of intermediate computations.

\subsubsection{Validation}

Model performance is evaluated on held-out test data with gesture samples using encrypted inference. Final predictions are decrypted only at authorized endpoints, and classification accuracy is computed in plaintext. This design ensures that neither raw motion features nor intermediate representations are exposed during training or inference, aligning with the system’s end-to-end privacy guarantees.

\section{Evaluation}
\subsection{Device Setup}
\label{sec:device_setup}

\begin{figure*}[!h]
    \centering
    \begin{subfigure}{0.325\textwidth}
        \centering
        \includegraphics[width=\textwidth]{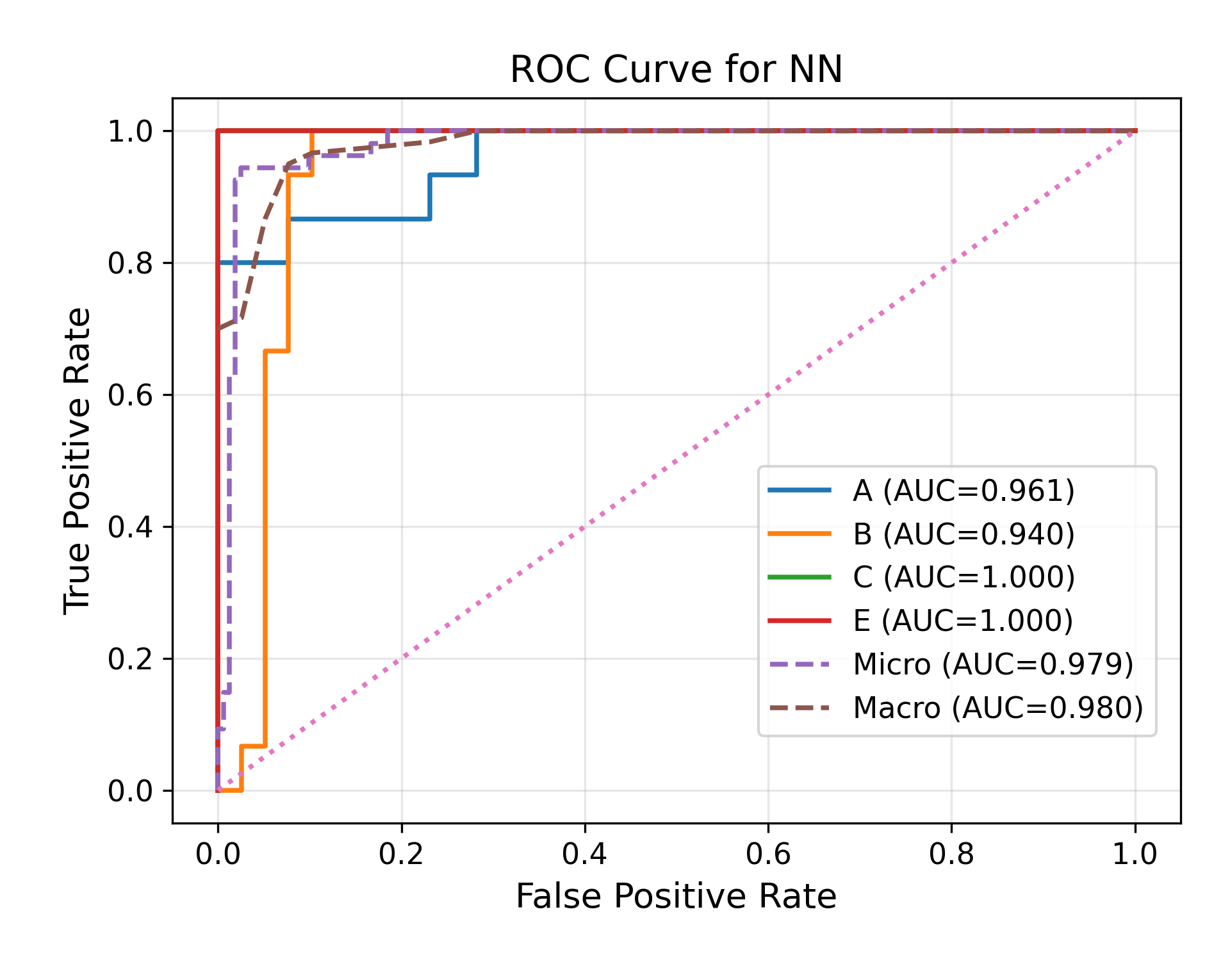}
        \caption{Desktop GPU (RTX 4090)}
    \end{subfigure}
    \hfill
    \begin{subfigure}{0.325\textwidth}
        \centering
        \includegraphics[width=\textwidth]{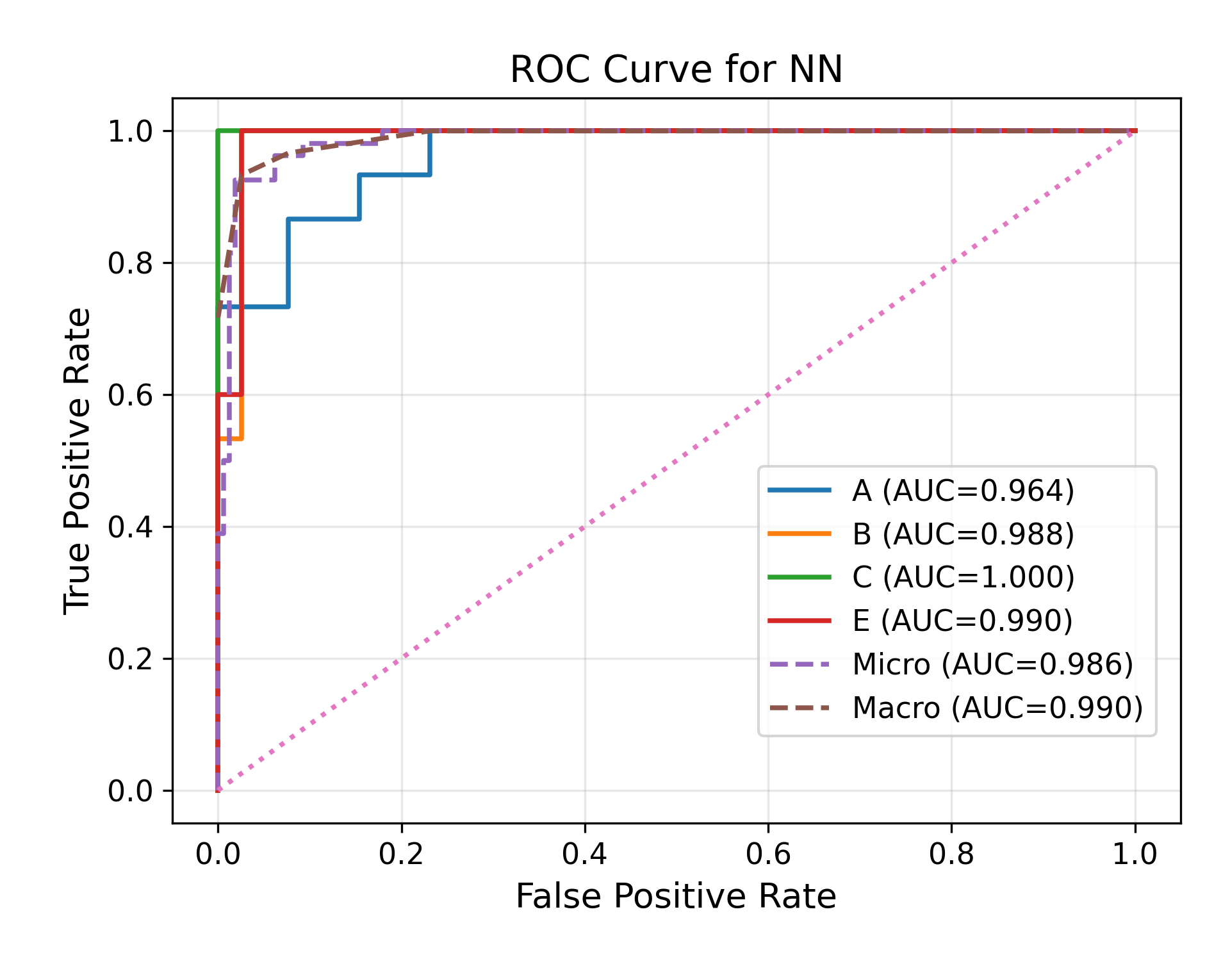}
        \caption{Nvidia Jetson Orin Nano}
    \end{subfigure}
    \hfill
    \begin{subfigure}{0.325\textwidth}
        \centering
        \includegraphics[width=\textwidth]{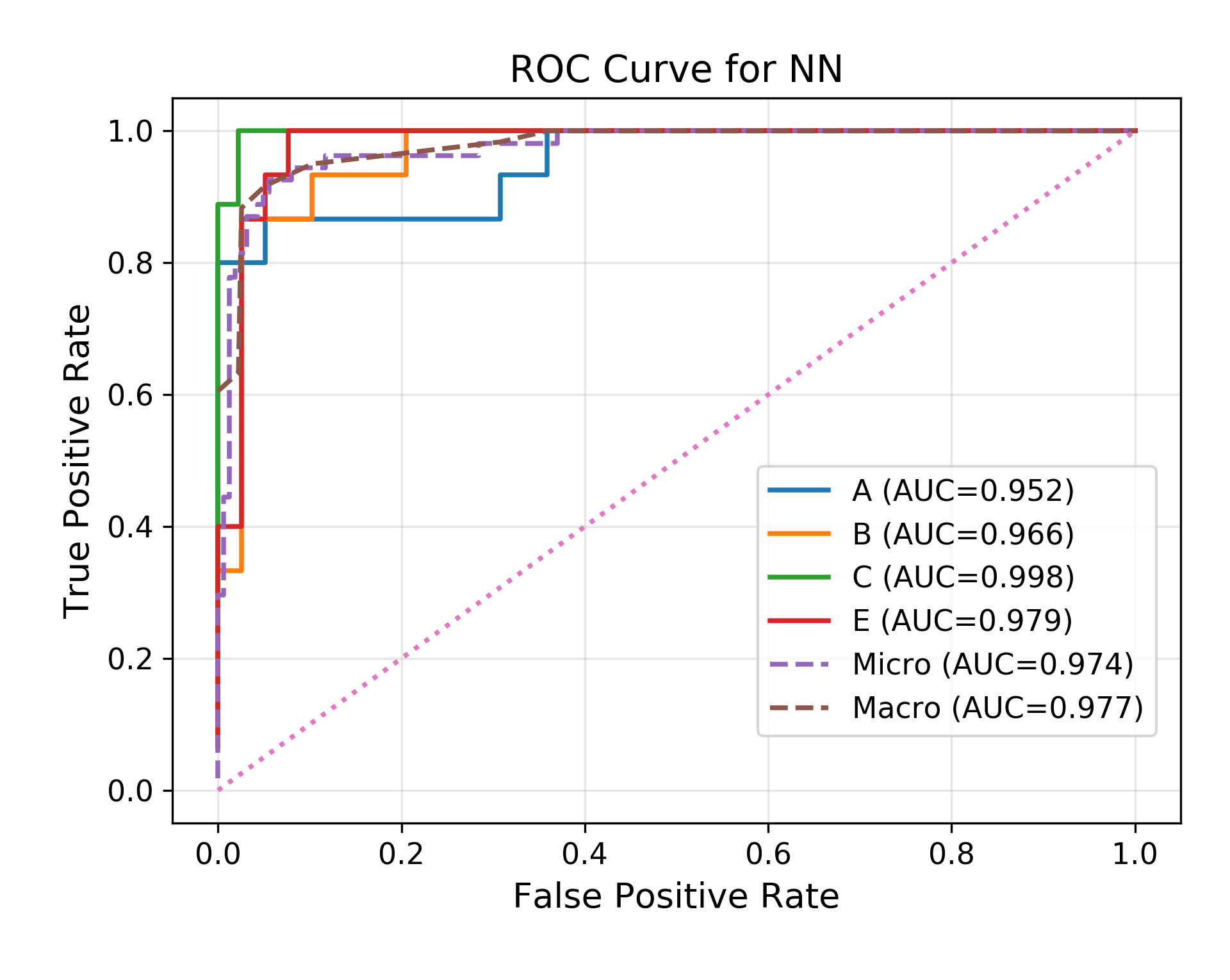}
        \caption{Nvidia Jetson Nano}
    \end{subfigure}
    \caption{ROC curve of general neural network trained on different devices}
    \label{roc-nn}
\end{figure*}

\begin{figure*}[!h]
    \centering
    \begin{subfigure}{0.325\textwidth}
        \centering
        \includegraphics[width=\textwidth]{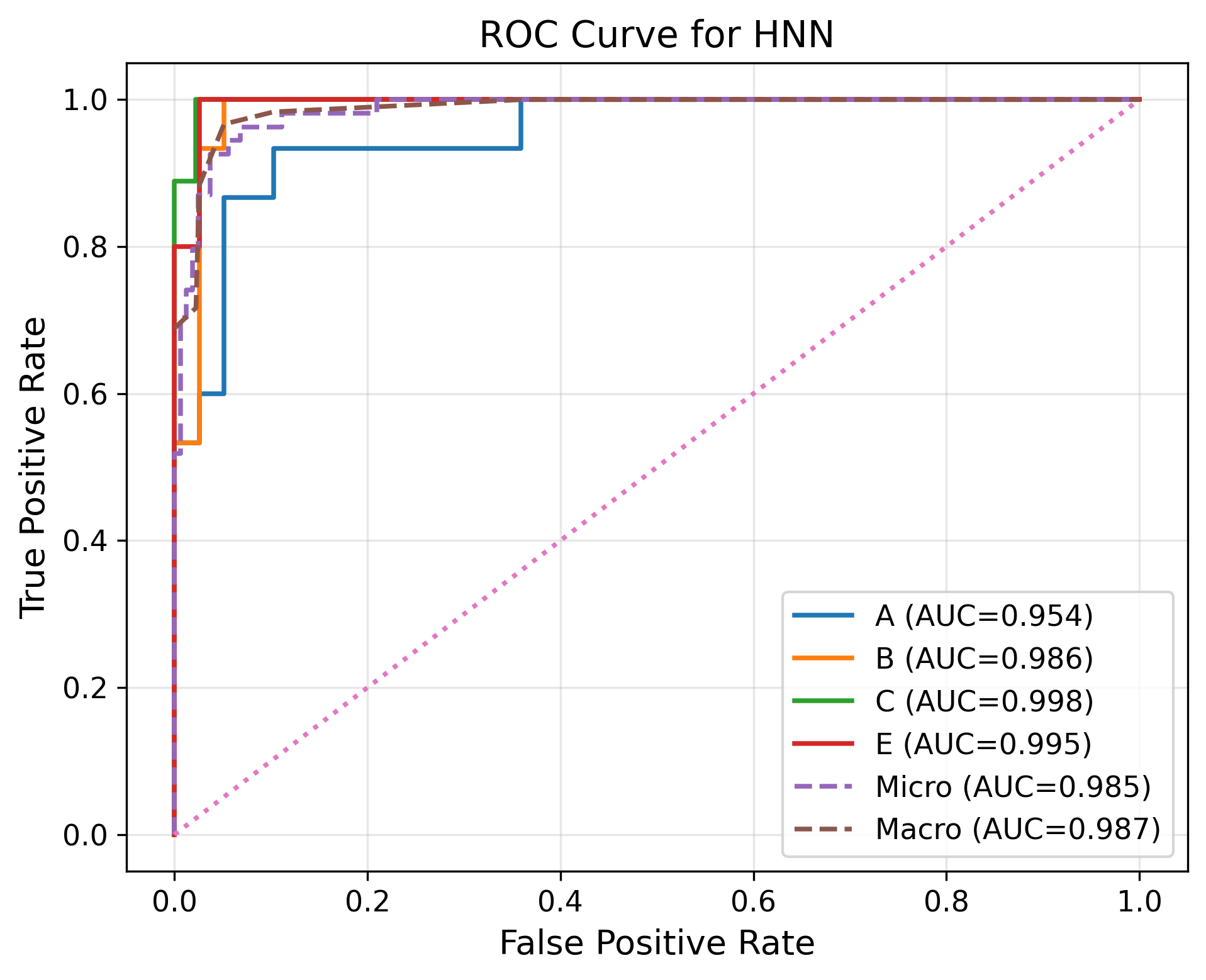}
        \caption{Desktop GPU (RTX 4090)}
    \end{subfigure}
    \hfill
    \begin{subfigure}{0.325\textwidth}
        \centering
        \includegraphics[width=\textwidth]{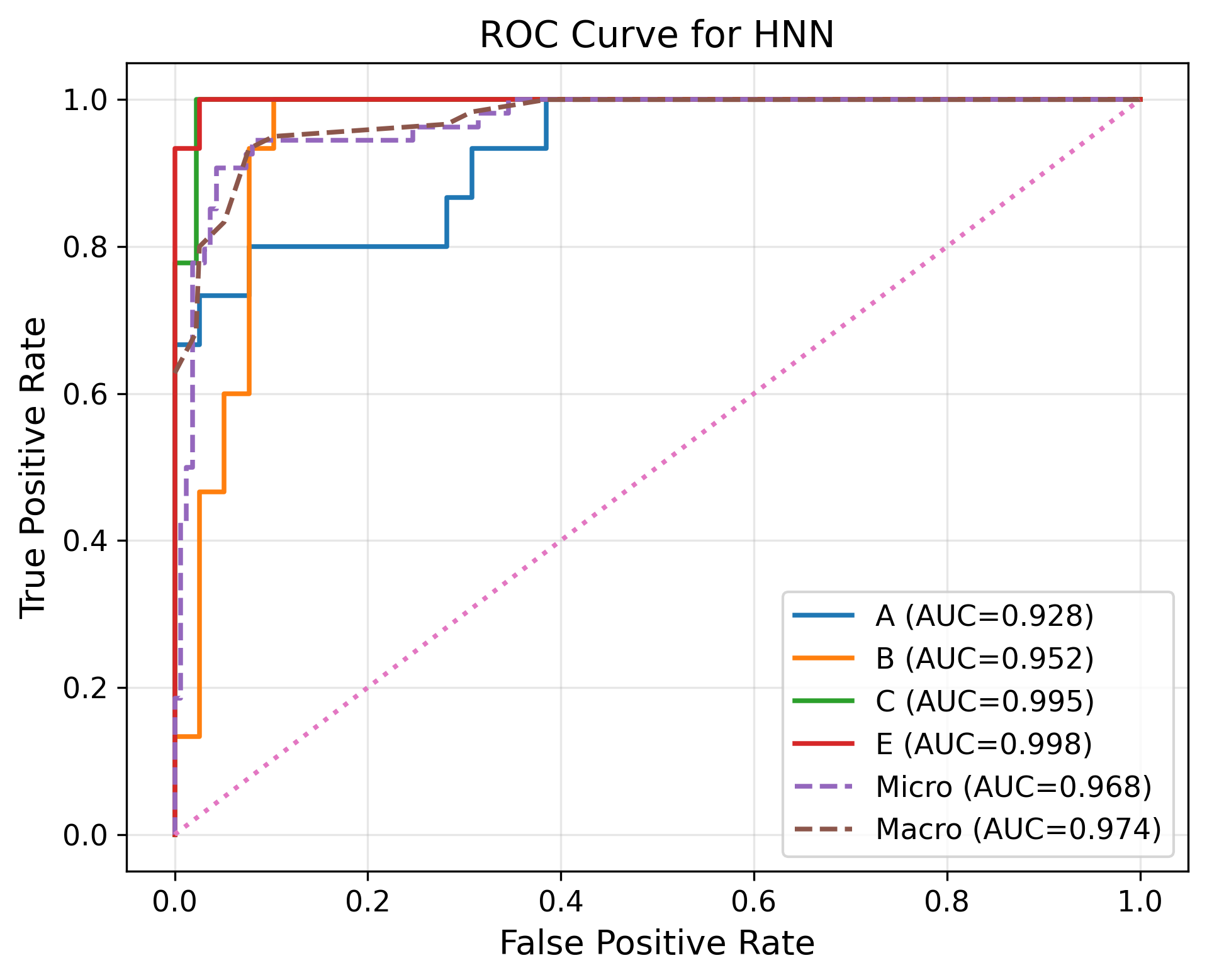}
        \caption{Nvidia Jetson Orin Nano}
    \end{subfigure}
    \hfill
    \begin{subfigure}{0.325\textwidth}
        \centering
        \includegraphics[width=\textwidth]{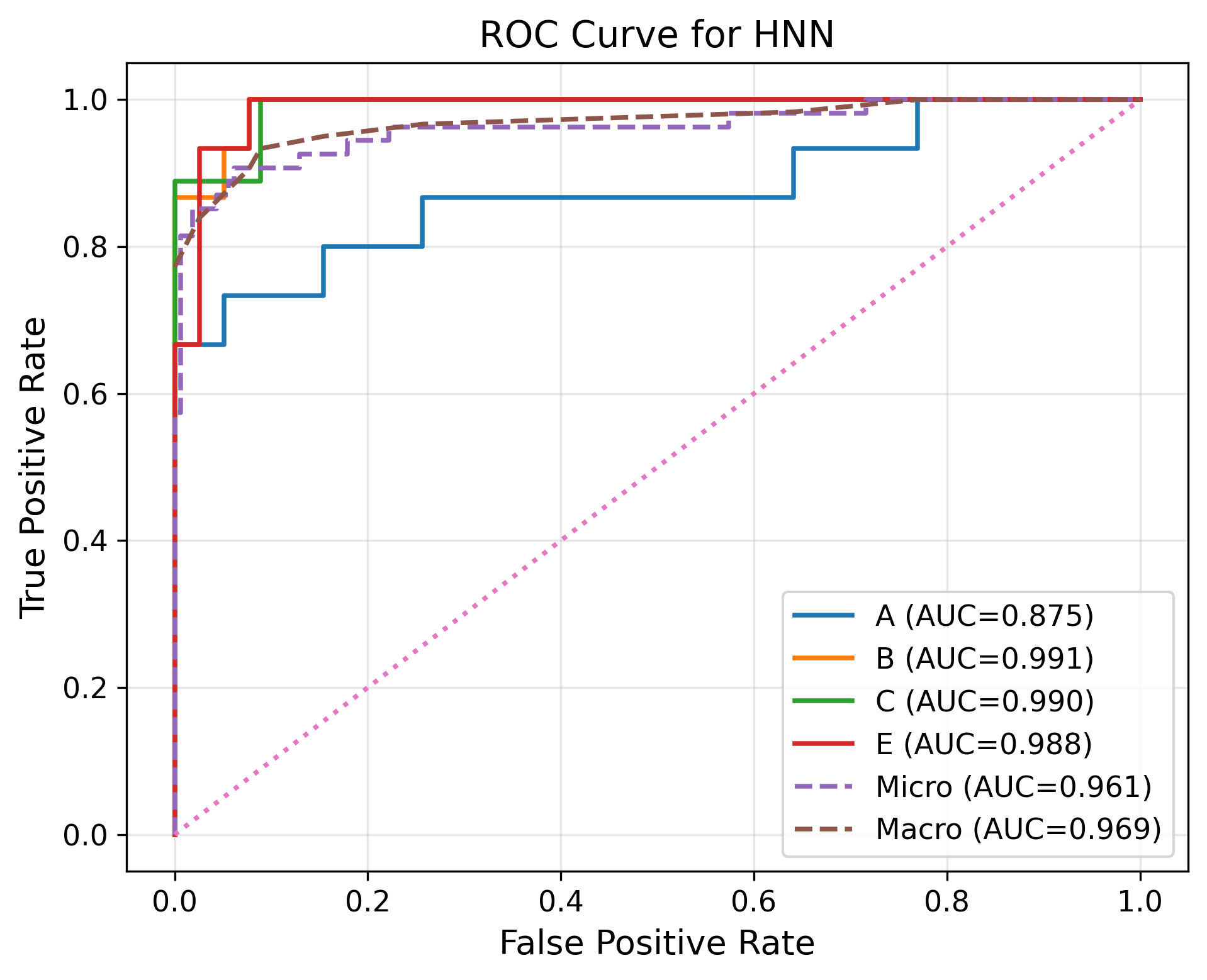}
        \caption{Nvidia Jetson Nano}
    \end{subfigure}
    \caption{ROC curve of homomorphic neural network trained on different devices}
    \label{roc-hnn}
\end{figure*}

Experiments were conducted on three heterogeneous platforms selected to span realistic deployment environments, from datacenter-class acceleration to severely resource-constrained edge hardware. The \emph{Lambda Vector One} system serves as a datacenter upper bound, equipped with an NVIDIA GeForce RTX~4090 (24~GB) and an AMD Ryzen~9~7950X CPU, establishing best-case training and inference performance under homomorphic computation \cite{lambda_vector_one}. The \emph{NVIDIA Jetson Orin Nano Developer Kit} represents a modern embedded AI platform with a 6-core Arm Cortex-A78AE CPU, an integrated Ampere GPU with 1024 CUDA cores, and 8~GB LPDDR5 memory, reflecting practical on-device inference scenarios such as assistive and mobile systems \cite{nvidia_jetson_orin_nano_datasheet,nvidia_jetson_orin_nano_brief}. Finally, the \emph{NVIDIA Jetson Nano Developer Kit} represents a legacy, highly resource-constrained edge platform with a quad-core Cortex-A57 CPU, a 128-core Maxwell GPU, and 2~GB LPDDR4 memory \cite{nvidia_jetson_nano_specs}; its inclusion enables stress-testing of the homomorphic pipeline under extreme compute and memory constraints, quantifying feasibility and performance degradation on low-end hardware. Together, these devices enable a controlled study of how homomorphic inference scales across datacenter, contemporary edge, and legacy embedded environments.

\begin{figure}[!h]
     \centering
         \includegraphics[width=0.5\textwidth]{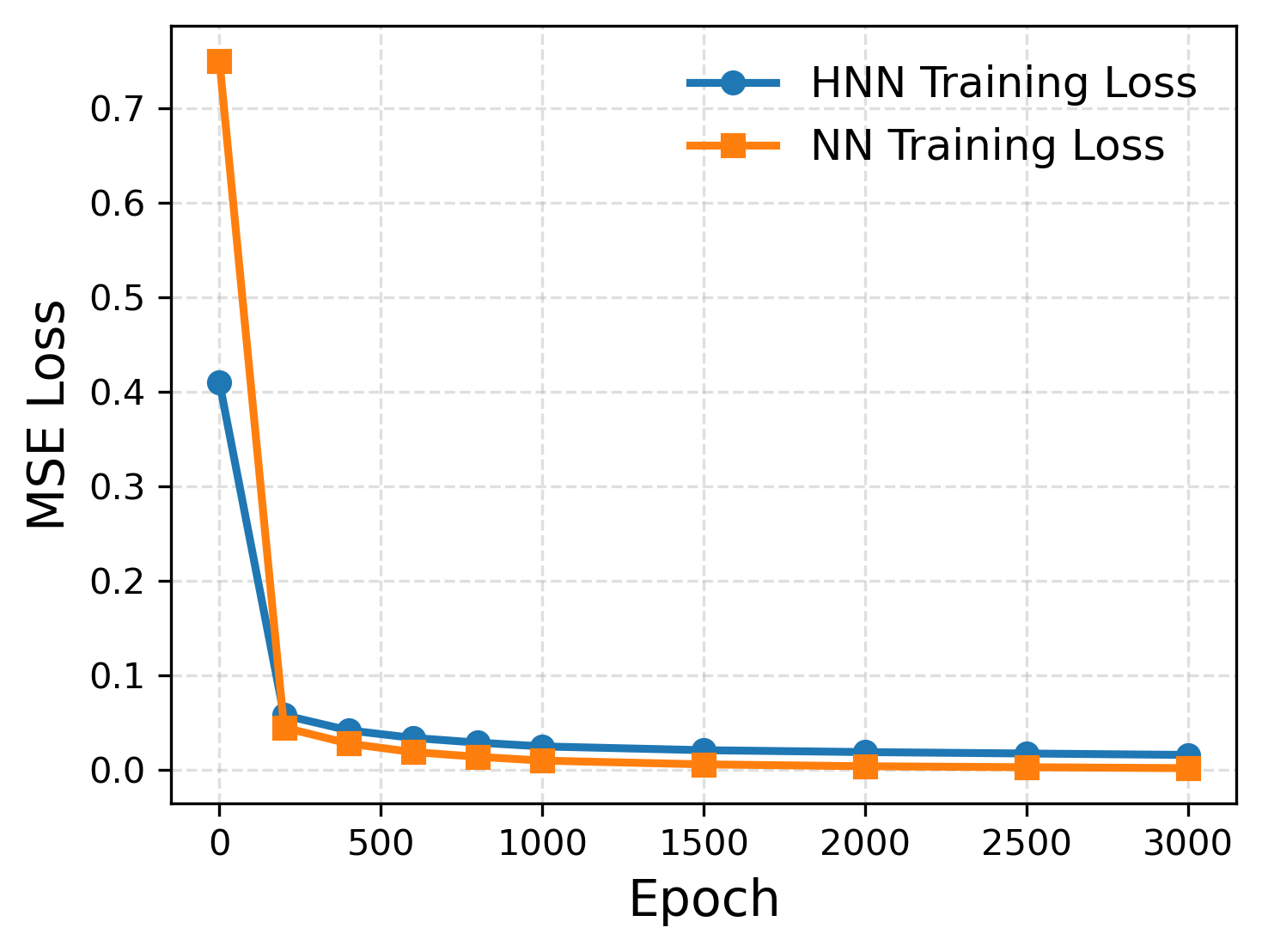}
        \caption{Comparison of training loss of General Neural Network and Homo Morphic Neural Network}
        \label{fig:training_loss}
\end{figure}

\begin{table*}[t]
\centering
\small
\resizebox{\textwidth}{!}{
\begin{tabular}{c|cccc|cccc}
\hline
\textbf{Device} &
\multicolumn{4}{|c|}{\textbf{NN}} &
\multicolumn{4}{|c}{\textbf{HNN}} \\ \cline{2-9}

\textbf{} &
\textbf{\textit{Train (s)}} &
\textbf{\textit{Lat. (ms)}} &
\textbf{\textit{Batch Lat. (ms)}} &
\textbf{\textit{Avg. Inf. (ms)}} &
\textbf{\textit{Train (s)}} &
\textbf{\textit{Lat. (ms)}} &
\textbf{\textit{Batch Lat. (ms)}} &
\textbf{\textit{Avg. Inf. (ms)}} \\

\hline \hline
Desktop GPU (RTX 4090) &
2.79 & 0.028 & 0.094 & 0.004 &
322.17 & 6.106 & 9.235 & 0.190 \\

Jetson Orin Nano &
65.03 & 0.305 & 2.258 & 0.020 &
1772.15 & 45.988 & 65.841 & 1.249 \\

Jetson Nano 2GB &
1107.04 & 0.388 & 1.419 & 0.304 &
4747.58 & 172.464 & 220.780 & 4.184 \\
\hline
\end{tabular}
}
\caption{Training and inference timing comparison for NN vs HNN. Latency corresponds to batch size 1; batch latency corresponds to batch size 54.}
\label{tab:timing_all}
\end{table*}

\begin{table*}[t]
\centering
\small
\resizebox{\textwidth}{!}{
\begin{tabular}{c|ccccc|ccccc}
\hline
\textbf{Device} &
\multicolumn{5}{|c|}{\textbf{NN}} &
\multicolumn{5}{|c}{\textbf{HNN}} \\ \cline{2-11}

\textbf{} &
\textbf{\textit{MSE}} &
\textbf{\textit{RMSE}} &
\textbf{\textit{Precision}} &
\textbf{\textit{Recall}} &
\textbf{\textit{F1}} &
\textbf{\textit{MSE}} &
\textbf{\textit{RMSE}} &
\textbf{\textit{Precision}} &
\textbf{\textit{Recall}} &
\textbf{\textit{F1}} \\

\hline \hline
Desktop GPU (RTX 4090) &
0.06543 & 0.25578 & 0.9507 & 0.9444 & 0.9430 &
0.07296 & 0.27011 & 0.9363 & 0.9259 & 0.9251 \\

Jetson Orin Nano &
0.07028 & 0.26510 & 0.9500 & 0.9444 & 0.9428 &
0.08043 & 0.28360 & 0.9333 & 0.9259 & 0.9222 \\

Jetson Nano 2GB &
0.06609 & 0.25709 & 0.9112 & 0.9074 & 0.9063 &
0.06990 & 0.26438 & 0.9178 & 0.9074 & 0.9045 \\
\hline
\end{tabular}
}
\caption{Output-space error and classification performance for NN vs HNN. Precision, recall, and F1 are weighted averages across classes.}
\label{tab:quality_all}
\end{table*}

\subsection{Training Dynamics and Convergence}
Figure~\ref{fig:training_loss} compares the training loss trajectories of the General Neural Network (NN) and the Homomorphic Neural Network (HNN). The General NN exhibits a steeper initial decrease in mean squared error (MSE), converging rapidly within the first few hundred epochs. In contrast, the HNN converges more gradually and stabilizes at a slightly higher asymptotic loss. This behavior is expected due to the restricted arithmetic and ciphertext noise growth inherent to homomorphic computation. Despite the slower convergence, the HNN remains numerically stable throughout training across all devices, with no divergence or oscillatory behavior observed. 

\begin{figure}[!h]
\centering
\begin{subfigure}{0.49\textwidth}
  \centering
  \includegraphics[width=0.75\linewidth]{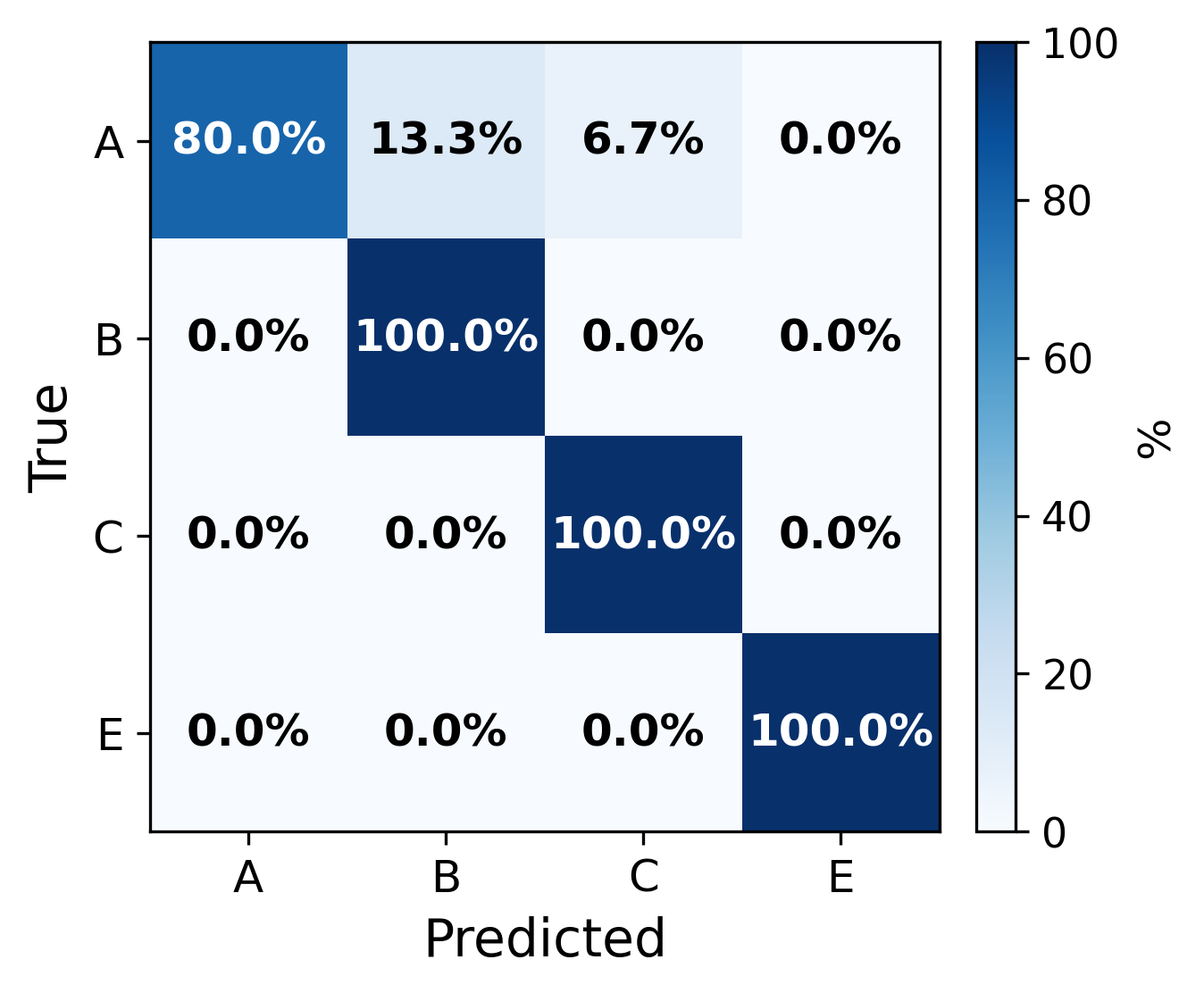}
  \caption{General Neural Network}
  \label{fig:cm-nn}
\end{subfigure}
\hfill
\begin{subfigure}{0.49\textwidth}
  \centering
  \includegraphics[width=0.75\linewidth]{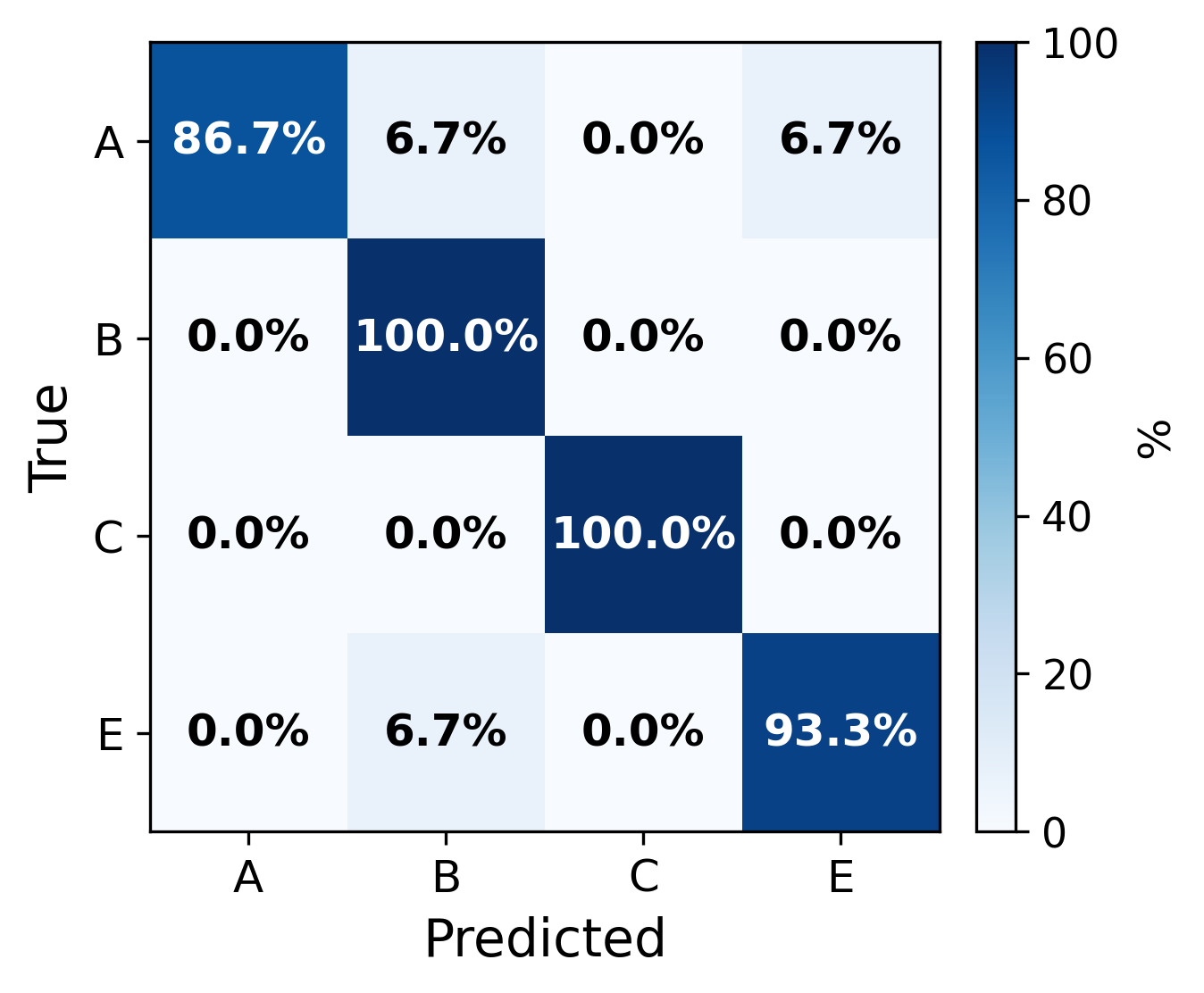}
  \caption{Homomorphic Neural Network}
  \label{fig:cm-hnn}
\end{subfigure}

\caption{Comparison of confusion matrices for the general and homomorphic neural networks.}
\label{fig:cm}
\end{figure}

Figures~\ref{roc-nn} and~\ref{roc-hnn} present ROC curves for the General NN and HNN, respectively, evaluated on the Desktop GPU (RTX~4090), Jetson Orin Nano, and Jetson Nano 2GB. Across all devices, both models achieve strong class separability, with micro- and macro-averaged AUC values consistently exceeding 0.96. On the RTX~4090, the General NN achieves near-optimal discrimination, with macro-AUC approaching 0.99, while the HNN achieves a macro-AUC of approximately 0.97. This modest reduction reflects the computational constraints of encrypted inference rather than a fundamental degradation in model expressiveness. Per-class ROC curves remain tightly clustered near the top-left corner for both models, indicating balanced class-wise performance. On the Jetson Orin Nano, ROC characteristics closely match those of the desktop GPU for both NN and HNN. This result demonstrates that reduced compute and memory budgets do not materially affect classification quality once training has converged. On the Jetson Nano 2GB, a larger performance gap is observed for certain classes, particularly under HNN inference; nevertheless, micro-AUC values remain above 0.96, confirming the viability of encrypted inference even on severely constrained legacy hardware.

\subsection{Classification performance}
Figure~\ref{fig:cm} presents normalized confusion matrices for both models. The General NN achieves near-perfect classification for classes B, C, and E, with minor confusion primarily concentrated in class A, and reaches a best overall classification accuracy of \textbf{94.44\%} on the Jetson Orin Nano platform. The HNN exhibits a similar pattern, with slightly increased misclassification for class A, most often confused with class B, while achieving a best accuracy of \textbf{92.59\%} on both the Desktop GPU (RTX~4090) and Jetson Orin Nano. This trend is consistent across devices and aligns with the ROC analysis, suggesting that errors arise from intrinsic class overlap rather than homomorphic noise. Importantly, no systematic bias toward false positives or false negatives is introduced by encrypted inference.


Table~\ref{tab:quality_all} reports output-space error (MSE/RMSE) and weighted classification metrics. Across all devices, the HNN incurs a modest increase in MSE and RMSE relative to the General NN, while precision, recall, and F1-score remain high; for example, on the RTX~4090 the HNN’s RMSE increases by approximately 0.014 with less than a two-point drop in F1-score. The consistent ordering of device performance across both models indicates that homomorphic computation preserves output fidelity without introducing device-specific instability.

\subsection{Runtime Performance}
Table~\ref{tab:timing_all} summarizes the computational overhead of privacy-preserving learning. Across all platforms, HNN training and inference are substantially slower than General NN due to encrypted arithmetic, with training on the RTX~4090 approximately two orders of magnitude slower and per-sample latency increasing from microseconds to milliseconds, while remaining within interactive bounds. On the Jetson Orin Nano, HNN inference reaches approximately 46~ms per sample at batch size~1 and is significantly amortized under batching, demonstrating feasibility on modern embedded platforms. The Jetson Nano~2GB represents a lower-bound stress test, where inference exceeds 170~ms per sample. 


The evaluation results directly validate the core design requirements of the proposed system. \textbf{Stealth} is supported by the measured inference latency on modern embedded platforms, where encrypted inference remains within interactive bounds and can be deployed entirely on-device, avoiding observable external communication. \textbf{Accuracy under encryption} is demonstrated by consistent classification performance exceeding 92\% across platforms, confirming that homomorphic inference preserves recognition fidelity despite operating exclusively on encrypted data. \textbf{Edge feasibility} is validated by successful deployment on the Jetson Orin Nano, which achieves practical latency and throughput while maintaining identical privacy guarantees to datacenter execution.




\section{Covert Feedback System}

\label{feedback}

The final stage of the communication pipeline delivers inferred gesture signals to the receiver in a covert and unobtrusive manner. To support reliable signal delivery across diverse operational contexts, we design and evaluate two complementary feedback mechanisms on commodity smartwatches: a haptic feedback channel and a visual feedback channel. Both mechanisms are designed to preserve discretion under casual observation while remaining usable in real time.

\subsection{Haptic Feedback.}
\begin{figure*}[ht]
  \centering
     \begin{subfigure}[b]{0.46\textwidth}
         \centering
         \includegraphics[width=0.85\textwidth]{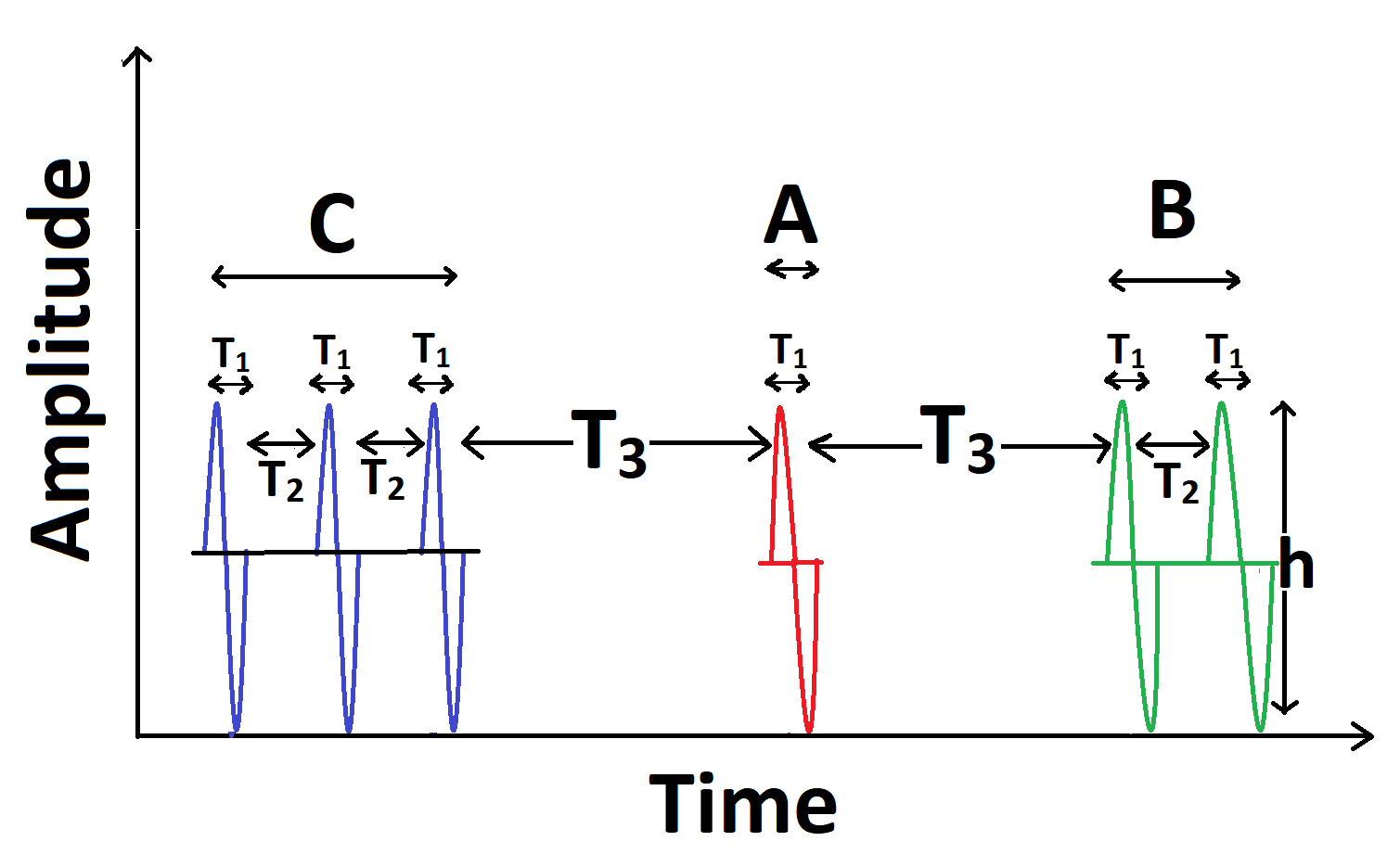}
         \caption{Vibration pattern representation.}
         \label{fig:haptic}
     \end{subfigure}
     \hfill
     \begin{subfigure}[b]{0.5\textwidth}
         \centering
         \includegraphics[width=\textwidth]{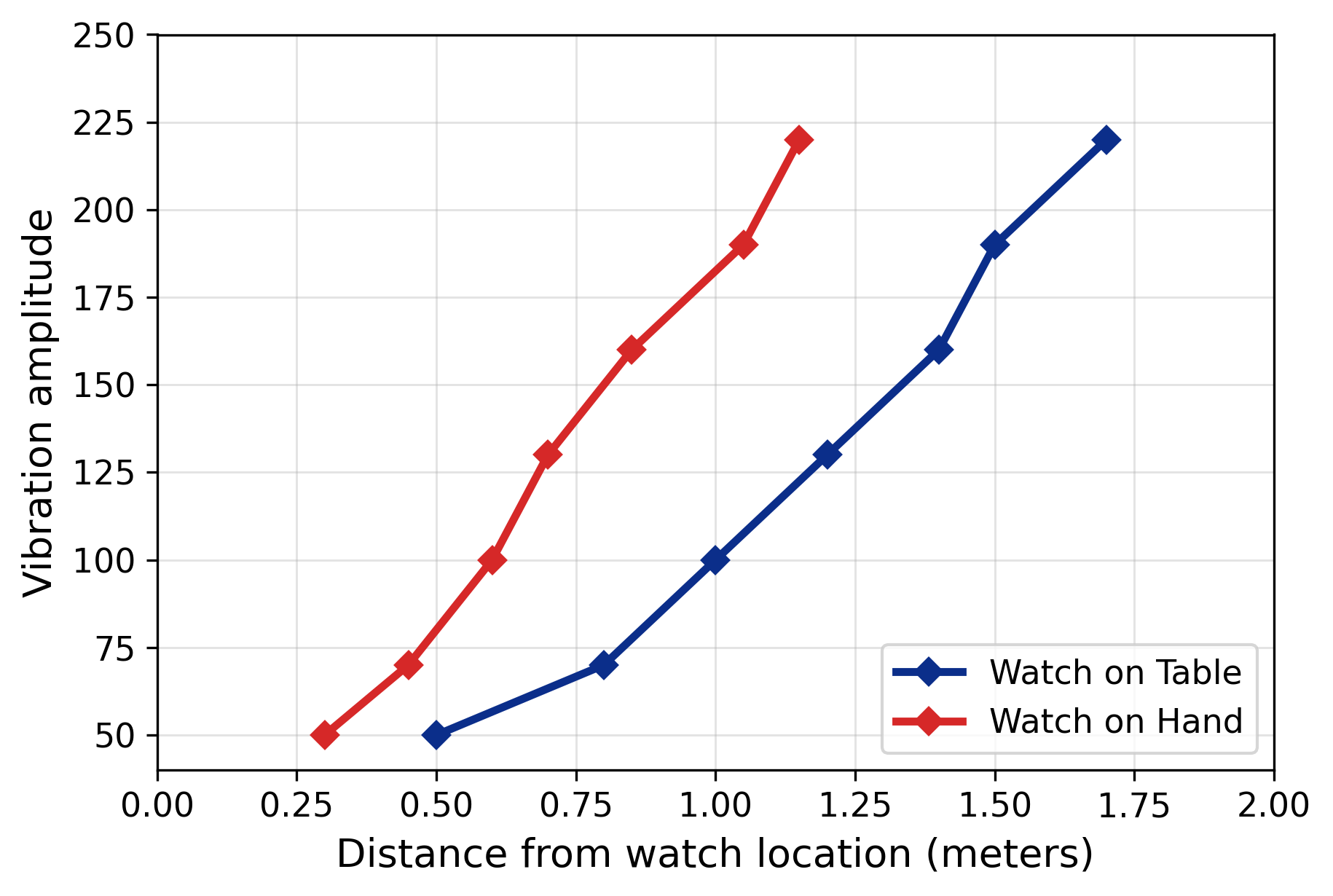}
         \caption{Distance vs vibrations in various amplitudes.}
         \label{fig-amp}
     \end{subfigure}
\caption{Illustration of the haptic-based system for covert communication feedback. Each vibration pattern represents a specific command or gesture.}
\label{prelil-2}
\end{figure*}

The haptic channel encodes inferred gesture signals as structured vibration patterns delivered through the smartwatch. Each gesture symbol is mapped to a distinct vibration sequence, enabling the receiver to decode signals through tactile perception without visual interaction. Figure~\ref{prelil-2} illustrates representative vibration patterns and their perceptual characteristics.

The haptic encoding (shown in figure \ref{fig:haptic}) is parameterized by four values: $T_1$, the duration of an individual vibration pulse; $T_2$, the interval between pulses within a sequence; $T_3$, the interval separating consecutive vibration sequences; and $h$, the vibration amplitude. These parameters are tuned to balance perceptibility for the wearer against the risk of audible or visible detection by nearby observers. In particular, $T_2$ and $T_3$ are selected to clearly distinguish pulses within a symbol from boundaries between symbols, while $h$ is chosen to remain below audibility thresholds in typical environments.

Figure \ref{fig-amp} evaluates the sensitivity of the haptic feedback channel to the vibration amplitude parameter $h$ by measuring the maximum distance at which vibrations become perceptible to nearby observers. Results compare two deployment configurations: the wearable worn on the user’s wrist and the device placed on a nearby surface. In both cases, perceptual range increases approximately linearly with amplitude. An amplitude of 70 units provided reliable perception by users while limiting observability to within approximately 0.5 meters when worn on the wrist. These results indicate that wrist-mounted haptic feedback offers stronger covertness in shared environments.

In scenarios where ambient conditions or proximity to others increase the risk of detection, the system allows the receiver to switch to the visual feedback channel.

\begin{figure}[ht]
  \centering
    \includegraphics[width=0.4\textwidth]{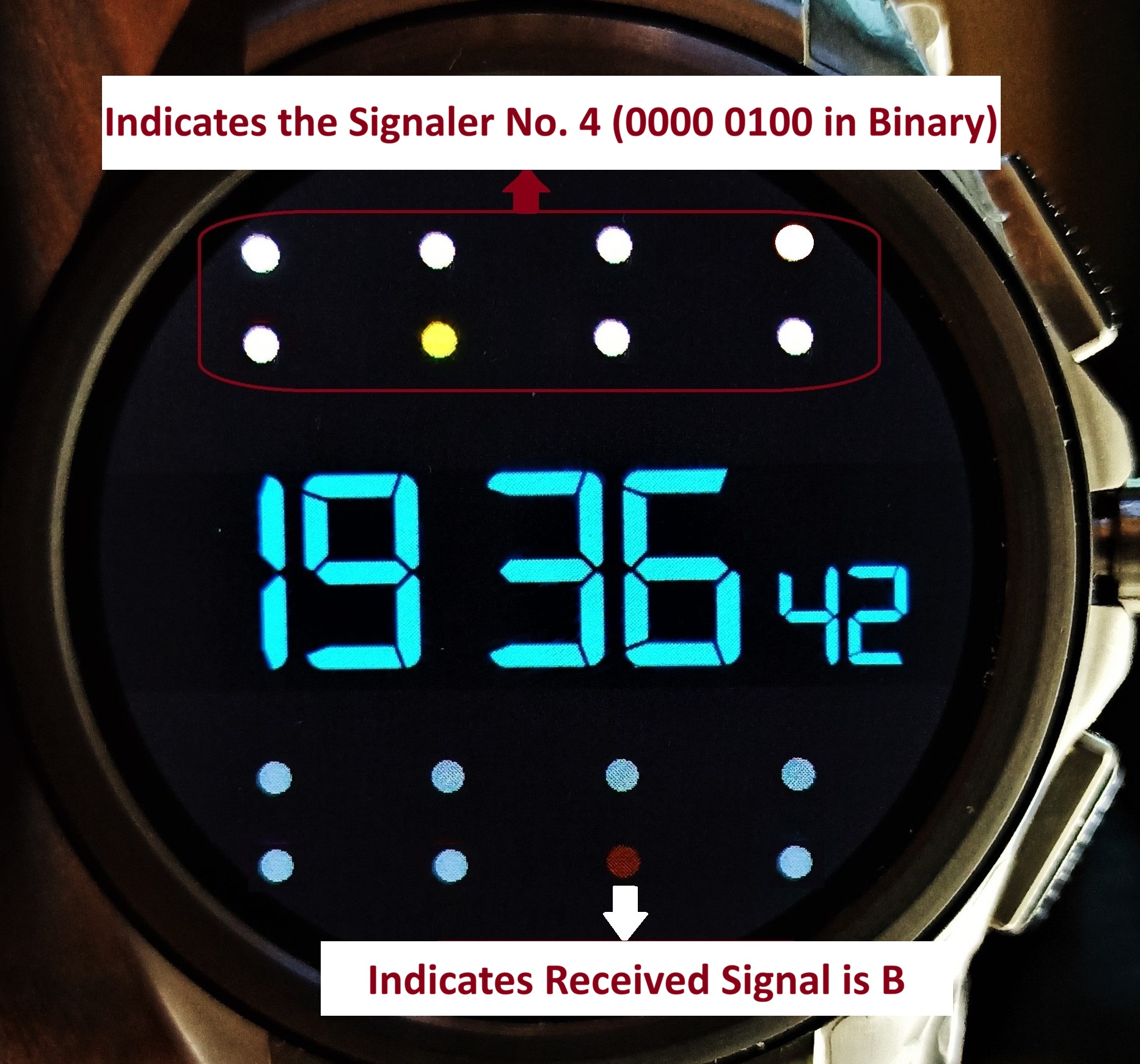}
\caption{Illustration of the visual feedback system.}
\label{fig:visual}
\end{figure}

\subsection{Visual Feedback System}
\label{sec:visual-feedback}

To support covert and reliable message delivery, the system provides a visual feedback mechanism that encodes inferred gesture symbols into an innocuous smartwatch display. Visual feedback is an alternative design solution for output channel other than the haptic feedback. The interface embeds signal information into a benign clock-like display that does not reveal its communication function under casual observation. As shown in Figure~\ref{fig:visual}, feedback is rendered using small, low-salience dots arranged in fixed regions of the screen. The lower region encodes the inferred gesture symbol, while the upper region optionally conveys sender metadata (e.g., identifier or session context).

\begin{table}[t]
\centering
\caption{Mapping between inferred gesture symbols and visual feedback elements}
\label{tab:gesture-visual-mapping}
\resizebox{0.85\textwidth}{!}{%
\begin{tabular}{c|c|c|c}
\hline 
\textbf{Gesture Symbol} & \textbf{Semantic Meaning} & \textbf{Dot Position (Grid)} & \textbf{Dot Color} \\
\hline \hline
A & Alert / Attention        & Bottom right, 1st dot  & Green  \\
B & Request / Action         & Bottom center, 2nd dot             & Red    \\
C & Acknowledge / Confirm    & Bottom center, 3rd dot             & Blue \\
E & Emergency / Abort        & Bottom left, 4th dot & Orange \\
\hline
\end{tabular}
}
\end{table}

Each gesture symbol is mapped to a unique visual code defined by a combination of dot position and color (Table~\ref{tab:gesture-visual-mapping}). This mapping is deterministic and consistent across sessions: for example, a red dot at the lower-center position corresponds to gesture~\texttt{B}, while a green dot at the lower-left position represents gesture~\texttt{A}. This design enables rapid recognition by trained users while remaining indistinguishable from decorative interface elements to external observers. Sender identifiers, when enabled, are encoded as a short binary sequence rendered as a row of white dots in the upper region of the display (Figure~\ref{fig:visual}). This allows receivers to distinguish among multiple authorized signalers without exposing explicit identifiers or textual content. Both gesture symbols and identifiers are displayed only after encrypted inference completes at the receiver side.


Both feedback mechanisms operate exclusively on decrypted outputs at the receiver device, ensuring that signal contents are never exposed to untrusted infrastructure. An external observer monitoring the device or communication channel cannot infer the presence, timing, or semantics of transmitted signals. Together, the haptic and visual channels provide complementary trade-offs between perceptibility, bandwidth, and environmental suitability, enabling flexible deployment across a range of covert communication scenarios.

\section{Conclusion}
This work demonstrates that covert, gesture-based communication on wearable devices can be made privacy-preserving by construction through end-to-end homomorphic inference. By ensuring that raw sensor data, intermediate features, and inference outputs remain encrypted throughout the entire pipeline, the proposed system eliminates opportunities for data leakage through untrusted infrastructure while preserving the usability and subtlety required for covert interaction. The system integrates commodity smartwatches and smartphones with homomorphic neural networks capable of operating on encrypted motion features, enabling secure inference on both cloud and edge platforms without altering the privacy model. Extensive evaluation across heterogeneous hardware shows that encrypted gesture inference retains high classification accuracy. In particular, results on modern embedded devices demonstrate that privacy-preserving inference can remain responsive in realistic deployment settings, while execution on legacy hardware establishes a lower bound on feasibility under extreme resource constraints. Rather than treating privacy as an add-on or relying on heuristic obfuscation, this work enforces confidentiality as a fundamental system property and explicitly quantifies the performance cost required to obtain it. 



\bibliographystyle{plain}
\bibliography{references}

\end{document}